\documentclass[sigconf]{acmart}
\usepackage{xcolor}
\definecolor{grey}{gray}{0.5}
\usepackage[T1]{fontenc}
\usepackage[utf8]{inputenc}
\usepackage[british]{babel}

\usepackage{algorithmic}
\usepackage{graphicx}
\usepackage{textcomp}
\usepackage{xcolor}
\usepackage{soul}
\usepackage{url}

\usepackage{booktabs}
\usepackage{array}
\usepackage{multirow}
\usepackage{tabularx}
\usepackage{threeparttable}
\usepackage{makecell}
\usepackage{adjustbox}
\usepackage{longtable}
\usepackage{supertabular}
\usepackage{multicol}
\usepackage{siunitx}
\usepackage{boldline}
\usepackage{colortbl}
\usepackage{balance}
\usepackage{caption}
\usepackage{subcaption}
\usepackage{enumitem}
\usepackage{chemformula}
\usepackage{mathrsfs}

\usepackage{tikz}
\usetikzlibrary{calc,arrows,decorations.markings,positioning,backgrounds,shapes,shapes.multipart}

\usepackage{listings}
\usepackage[skins,breakable]{tcolorbox}
\usepackage{xcolor}

\newcommand{\up}[1]{\textcolor{green!60!black}{\scriptsize $(\uparrow #1)$}}

\usepackage{amssymb}
\usepackage{colortbl}
\usepackage{booktabs}
\usepackage{graphicx}
\usepackage{colortbl}

\definecolor{pastelblue}{HTML}{aec6cf}
\definecolor{grayframe}{HTML}{d8d8d8}
\definecolor{pastelgreen}{HTML}{77dd77}
\definecolor{whiteframe}{HTML}{ffffff}
\definecolor{pastelpink}{HTML}{ffb6c1}
\definecolor{creamframe}{HTML}{ffffcc}
\definecolor{lightlavender}{HTML}{e6e6fa}
\definecolor{pastelgray}{HTML}{cfcfc4}
\definecolor{pastelyellow}{HTML}{fdfd96}
\definecolor{lightgrayframe}{HTML}{f0f0f0}
\definecolor{lighterpastelblue}{HTML}{cde0e5}
\definecolor{lighterlavender}{HTML}{f3f3fd}
\definecolor{verylightpastelblue}{HTML}{e0f0f5}
\definecolor{verylightpastelgreen}{HTML}{a3e4a7}
\definecolor{verylightwhiteframe}{HTML}{ffffff}
\definecolor{verylightpastelpink}{HTML}{ffdce0}
\definecolor{verylightcreamframe}{HTML}{ffffe6}
\definecolor{verylightlavender}{HTML}{fbfaff}
\definecolor{verylightpastelgray}{HTML}{f0f0f0}
\definecolor{verylightpastelyellow}{HTML}{ffffc2}
\definecolor{verylightlightgrayframe}{HTML}{f5f5f5}

\definecolor{burgundy}{RGB}{128,0,32}
\definecolor{lightburgundy}{RGB}{171,76,98}
\definecolor{darkgreen}{RGB}{0,100,0}
\definecolor{lightblue}{RGB}{173,216,230}
\definecolor{EMP}{HTML}{77DD77}
\definecolor{NOR}{HTML}{06500C}

\lstdefinestyle{SQLStyle}{
    language=SQL,
    basicstyle={\fontsize{6}{6}\ttfamily\selectfont},
    breaklines=true,
    breakatwhitespace=true,
    showstringspaces=false,
    keywordstyle=\color{blue},
    stringstyle=\color{lightburgundy},
    commentstyle=\color{green!60!black},
    numbers=none,
    tabsize=2,
    showtabs=false,
    frame=none,
    xleftmargin=0pt,
    aboveskip=5pt,
    belowskip=5pt
}

\lstdefinestyle{VerilogStyle}{
    language=Verilog,
    basicstyle=\ttfamily\fontsize{5}{5}\selectfont,
    keywordstyle=\color{darkgreen}\bfseries,
    commentstyle=\color{gray},
    stringstyle=\color{red},
    numbers=left,
    numberstyle=\tiny,
    stepnumber=1,
    numbersep=5pt,
    xleftmargin=10pt,
    framexleftmargin=10pt,
    breaklines=true,
    breakatwhitespace=false,
    showspaces=false,
    showstringspaces=false,
    showtabs=false,
    tabsize=2,
    morekeywords=[1]{module,endmodule,input,output,assign},
}

\lstdefinestyle{CypherStyle}{
    language=SQL,
    basicstyle={\fontsize{5}{5}\ttfamily\selectfont},
    breaklines=true,
    breakatwhitespace=true,
    showstringspaces=false,
    keywordstyle=\color{blue},
    stringstyle=\color{black},
    commentstyle=\color{green!60!black},
    numbers=none,
    tabsize=2,
    showtabs=false,
    frame=none,
    xleftmargin=0pt,
    aboveskip=5pt,
    belowskip=5pt,
    morekeywords={MATCH, RETURN, CREATE, MERGE, WHERE, DELETE, WITH, OPTIONAL, DETACH, SET, LIMIT, ORDER, BY, SKIP, UNWIND, CALL},
    morestring=[b]",
    moredelim=[s][\color{lightburgundy}]{-}{->},
    moredelim=[s][\color{lightburgundy}]{<}{-},
    moredelim=[s][\color{lightburgundy}]{-[}{]-},
}

\lstdefinelanguage{json}{
  basicstyle=\fontsize{6}{6}\selectfont\ttfamily,
  showstringspaces=false,
  breaklines=true,
  breakatwhitespace=true,
  literate=
    *{0}{{{\color{blue}0}}}{1}
     {1}{{{\color{blue}1}}}{1}
     {2}{{{\color{blue}2}}}{1}
     {3}{{{\color{blue}3}}}{1}
     {4}{{{\color{blue}4}}}{1}
     {5}{{{\color{blue}5}}}{1}
     {6}{{{\color{blue}6}}}{1}
     {7}{{{\color{blue}7}}}{1}
     {8}{{{\color{blue}8}}}{1}
     {9}{{{\color{blue}9}}}{1}
     {:}{{{\color{red}:}}}{1}
     {,}{{{\color{red},}}}{1}
     {\{}{{{\color{brown}\{}}}{1}
     {\}}{{{\color{brown}\}}}}{1}
     [{{{\color{brown}[}}}{1}
     ]{{{\color{brown}]}}}{1},
}

\tcbset{
  covrbox/.style={
    colback=white,
    colframe=black,
    width=\linewidth,
    boxrule=0.2mm,
    arc=1mm,
    left=3pt,
    right=3pt,
    top=2pt,
    bottom=1pt,
    before skip=0.6pt,
    after skip=0.1pt,
    fontupper={\fontsize{5}{6}\selectfont},
    fonttitle=\bfseries\fontsize{6}{6}\selectfont,
    coltitle=black,
    colbacktitle=cyan!15!blue!10
  }
}

\def\BibTeX{{\rm B\kern-.05em{\sc i\kern-.025em b}\kern-.08em
    T\kern-.1667em\lower.7ex\hbox{E}\kern-.125emX}}

\copyrightyear{2026}
\acmYear{2026}
\setcopyright{cc}
\setcctype{by}
\acmConference[MLCAD '26]{2026 ACM/IEEE International Symposium on Machine Learning for CAD}{September 07--09, 2026}{Jeju Island, Republic of Korea}
\acmBooktitle{2026 ACM/IEEE International Symposium on Machine Learning for CAD (MLCAD '26), September 07--09, 2026, Jeju Island, Republic of Korea}
\acmDOI{10.1145/3831599.3840347}
\acmISBN{979-8-4007-2878-5/2026/09}

\begin{document}

\title{CovR: Coverage-Aware Hardware Verification via Reasoning-Guided Reinforcement Learning}

\author{Manar Abdelatty}
\email{manar_abdelatty@brown.edu}
\affiliation{%
  \department{School of Engineering}
  \institution{Brown University}
  \city{Providence}
  \state{RI}
  \country{USA}
}

\author{Maryam Nouh}
\email{maryam_nouh@brown.edu}
\affiliation{%
  \department{School of Engineering}
  \institution{Brown University}
  \city{Providence}
  \state{RI}
  \country{USA}
}

\author{Sherief Reda}
\email{sherief_reda@brown.edu}
\affiliation{%
  \department{School of Engineering}
  \institution{Brown University}
  \city{Providence}  \state{RI}
  \country{USA}
}

\renewcommand{\shortauthors}{Trovato et al.}

\begin{abstract}
Design verification remains one of the most resource-intensive stages of hardware development, often consuming up to 70\% of the total design effort. While recent work has explored using Large Language Models (LLMs) to automate testbench generation, most existing approaches focus narrowly on functional correctness, overlooking the critical aspect of coverage quality. To bridge this gap, we present CovR, an agentic framework for automated testbench generation that combines self-reflection loops with simulation-based feedback to maximize coverage. Using this pipeline, we construct a large-scale dataset of  $16,514$ natural specification–RTL–reasoning–testbench tuples with a strong teacher model, enabling coverage-aware supervision. Building on this, we propose a reinforcement learning (RL) framework tailored for coverage-driven testbench generation, leveraging tool-derived rewards from simulation and coverage feedback to optimize a student model. Experimental results show that the CovR finetuned model achieves 93.81\% cov@10 on VerilogEval and RTLLM V2.0, and 87.76\% cov@10 on CVDP, outperforming state-of-the-art approaches by 7.97\% and 3.59\%, respectively. Furthermore, deploying the finetuned model back into the agentic refinement pipeline further improves cov@10 to 94.27\% on VerilogEval and RTLLM V2.0 and 91.39\% on CVDP. Moreover, when integrated as a plug-in stimulus engine for full verification workflows, \emph{CovR} improves coverage by 18.95\% and mutation detection score by 1.19\%, while revealing 4.46\% undetected failures, highlighting the importance of optimizing for coverage in LLM-based hardware verification.

\end{abstract}


\keywords{Large Language Model, LLM, Hardware Verification, Code Coverage, Testbench Generation, Reasoning}
  

\maketitle

\section{Introduction}

Design verification is a fundamental stage in the hardware design workflow, ensuring that a Register Transfer Level (RTL) implementation faithfully realizes its intended specification~\cite{bergeron2000writing, spear2012systemverilog}. This process typically consists of two complementary components. First, a \emph{Functional Reference Model} (FRM), or an equivalent specification-derived oracle, is constructed to capture the expected behavior of the design~\cite{huang2005principles}. Second, test stimuli are generated to exercise the RTL across diverse scenarios, while validating its outputs against the oracle using self-checking mechanisms such as assertions~\cite{ieee1800}. The effectiveness of this process is quantified along two dimensions: \emph{functional coverage}, which measures how well intended behaviors are exercised~\cite{mehta2020sva, piziali2004functional, fine2005coverage}, and  \emph{code coverage}, which evaluates how thoroughly the RTL implementation, including its lines, conditions, states, and branches, is explored during simulation~\cite{Wang1995practical}. Despite advances in design automation, verification can consume up to 70\% of the hardware development effort~\cite{carter2007metric}. This motivates the need for intelligent automation to accelerate verification and improve productivity.

\begin{figure}[!t]
    \centering
     \begin{subfigure}{1\linewidth}
      \captionsetup{skip=1pt} 
        \centering
    \includegraphics[width=\linewidth]{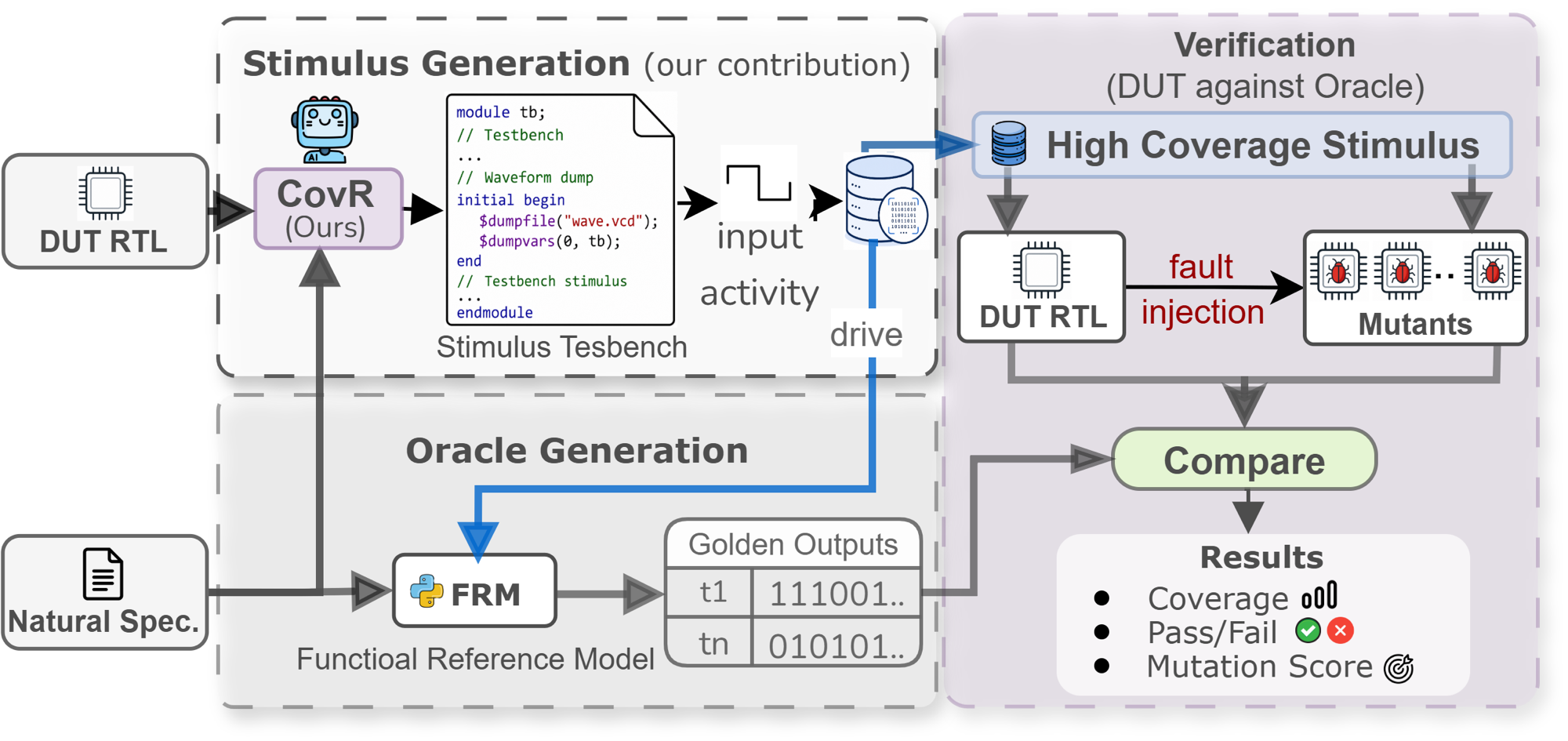}
        \centering
        \label{fig:cov_plug}
    \end{subfigure}
    \vspace{-4em}
\caption{Overview of \emph{CovR} within a full RTL verification workflow. \emph{CovR} generates high-coverage stimuli that exercise the DUT, oracle, and injected RTL mutants to improve coverage, mutation detection, and bug-finding effectiveness.}
\vspace{-1.8em}
\label{fig:covr_plugin}
\end{figure}

Recent work leverages Large Language Models (LLMs) to automate verification tasks, including the generation of testbenches~\cite{qiu2025correctbench, qiu2024autobench}, test plans~\cite{kochar2026grpo}, Functional Reference Models (FRMs)~\cite{zhao2025pro}, and assertions~\cite{yan2025assertllm,Zhang2025exploring}. While effective at producing executable verification artifacts, these approaches often overlook the diversity of generated stimuli, leading to limited coverage and reduced ability to uncover corner-case bugs. Coverage-guided methods address this by optimizing for coverage via iterative feedback~\cite{zhang2025llm4dv}, curriculum-based finetuning~\cite{zhang2026llm4cov}, preference optimization~\cite{nadimi2025tb}, and reinforcement learning~\cite{park2025llm4teststimuli}. However, they exhibit three key limitations: 
(i) methods such as LLM4DV~\cite{zhang2025llm4dv} and the RL-based approach in~\cite{park2025llm4teststimuli} represent stimuli as structured or binary sequences, limiting their expressiveness compared to hardware description languages such as Verilog.
(ii) State-of-the-art approaches such as LLM4COV~\cite{zhang2026llm4cov} rely on curriculum-based supervised finetuning over offline training data, rather than directly optimizing stimulus generation through online interaction with coverage feedback during training. (iii) Agentic pipelines in LLM4DV~\cite{zhang2025llm4dv} lack self-reflection~\cite{zhang2025agentic, shinn2023reflexion}, limiting their ability to accumulate knowledge across iterations and converge to high-coverage solutions.

To address these limitations, we introduce \emph{CovR}\footnote{https://github.com/orgs/scale-lab/CovR}, an agentic framework for high-coverage stimulus generation. \emph{CovR} iteratively refines test stimuli using simulation feedback and trains a reasoning-enhanced LLM through supervised finetuning and reinforcement learning with rewards derived from simulation validity and coverage reports. Fig.~\ref{fig:covr_plugin} illustrates how \emph{CovR} integrates into full verification workflows as a plug-in stimulus engine, generating high-coverage stimuli that drive design-under-test (DUT), oracle, and design mutants to improve coverage closure and bug detection.

Our contributions are as follows:
\begin{itemize}[topsep=2pt,itemsep=2pt,parsep=0pt,partopsep=0pt]
\item We develop an agentic pipeline that leverages simulation feedback and self-reflection to iteratively refine testbenches toward high coverage, enabling automatic generation of high-quality synthetic data via a strong \emph{teacher} model.

\item Using the proposed pipeline, we construct a large-scale dataset for high-coverage testbench generation, comprising $16{,}514$ specification–RTL–testbench–reasoning tuples. 

\item We formulate test stimulus generation as a coverage optimization problem and propose a learning framework that combines supervised finetuning on reasoning-augmented data with reinforcement learning using tool-derived rewards.

\item We train a \emph{student} model with our learning framework to approximate the \emph{teacher} pipeline, improving average \emph{cov@1} by 44.29\% over the base model and surpassing the teacher by 7.3\% on average across VerilogEval, RTLLM, and CVDP.

\item We deploy the finetuned model into the \emph{CovR} self-refinement agentic pipeline, further improving average \emph{cov@1} by  6.68\% compared to direct inference.  

\item We integrate \emph{CovR} into full verification workflows~\cite{zhao2025pro,qiu2025correctbench} as a plug-in stimulus engine, improving coverage by 18.95\%, mutation detection score by 1.19\% on average, while revealing 4.46\% more previously undetected failures.

\end{itemize}
This paper is organized as follows. Section ~\ref{related_work} discusses related work. Section ~\ref{benchmark} presents the \emph{CovR} self-refinement workflow, synthetic dataset generation, the finetuning process. Section ~\ref{experimental_results} presents our experimental results. Finally, section ~\ref{conclusion} concludes the paper. 

\section{Related Work}
\label{related_work}
Recent work applies LLMs to hardware verification through two main directions: testbench generation frameworks and coverage-guided stimulus generation, differing in their use of representations, learning strategies, and feedback mechanisms.

\textbf{Testbench Generation Frameworks} AutoBench~\cite{qiu2024autobench} and CorrectBench~\cite{qiu2025correctbench} generate assertion-based testbenches by constructing a Python reference model as an oracle, followed by iterative refinement to improve bug detection. However, both incur high computational cost due to repeated prompting and lack of finetuning. PRO-V-R1~\cite{zhao2025pro} extends this paradigm with a multi-agent  framework that incorporates supervised finetuning and reinforcement learning to improve reasoning and tool use, reducing reliance on iterative prompting. Similarly, the work in~\cite{kochar2026grpo} improves test plan generation via reinforcement learning. Despite these advances, existing methods primarily target functional correctness and bug detection, with limited emphasis on test stimuli coverage.

\textbf{Coverage-Guided Generation} Coverage-driven approaches use simulation feedback to improve stimulus quality. LLM4DV ~\cite{zhang2025llm4dv} targets \emph{functional coverage} through an iterative loop between a Python-based test generator and a simulation backend (e.g., Verilator~\cite{verilator}), using manually specified coverage bins implemented in cocotb~\cite{cocotb}. However, it relies on repeated prompting without self-reflection~\cite{shinn2023reflexion,zhang2025agentic} or model finetuning, requiring many iterations to achieve acceptable coverage. 
Later work~\cite{park2025llm4teststimuli} introduces supervised and reinforcement learning for coverage optimization, but represents stimuli as table-structured binary matrices. As a result, it scales poorly with input dimensionality, lacks semantic structure, and cannot effectively capture temporal behaviors, limiting generalization to complex designs. The work in~\cite{nadimi2025tb} further explores finetuning via Direct Preference Optimization (DPO)~\cite{rafailov2023direct} to favor high-coverage testbenches, but reliance on static preference data limits exploration compared to tool-interactive or reinforcement learning-based methods. LLM4COV ~\cite{zhang2026llm4cov} advances finetuning with a curriculum-based strategy that gradually increases coverage complexity during supervised finetuning. However, it is limited to SFT and does not leverage reinforcement learning to explore diverse high-coverage behaviors through interactive tool feedback.



\begin{table}[!t]
\captionsetup{skip=2pt}
\fontsize{7}{8}\selectfont
\centering
\caption{Prior work leverages coverage feedback but rarely integrates reasoning with reinforcement learning. \emph{CovR} unifies these components to produce high-coverage testbenches.}\label{tab:lit_review}
\setlength{\tabcolsep}{3pt}
\renewcommand{\arraystretch}{0.85}

\resizebox{\columnwidth}{!}{

\begin{tabular}{l|cccll}
\toprule
\textbf{Approach} 
& \shortstack{\textbf{Coverage}\\\textbf{Feedback}} 
& \textbf{Agentic} 
& \textbf{Reasoning} 
& \textbf{Learning} 
& \shortstack{\textbf{Primary}\\\textbf{Output}} \\
\midrule

AutoBench~\cite{qiu2024autobench} 
& $\times$ 
& $\checkmark$ 
& $\times$ 
& None 
& FRM + Stimuli Testbench \\

CorrectBench~\cite{qiu2025correctbench} 
& $\times$ 
& $\checkmark$ 
& $\times$ 
& None 
& FRM + Stimuli Testbench \\

PRO-V-R1~\cite{zhao2025pro}
& $\times$ 
& $\checkmark$ 
& $\checkmark$ 
& SFT-W/Reason + RL 
& FRM + Test Stimuli\\

Testplan Gen.~\cite{kochar2026grpo}
& $\times$ 
& $\checkmark$ 
& $\checkmark$ 
& SFT-W/Reason + RL  
& Test-plan \\

LLM4DV~\cite{zhang2025llm4dv}
& $\checkmark$ 
& $\checkmark$ 
& $\times$ 
& None 
& Test stimuli \\

Test stimuli gen.~\cite{park2025llm4teststimuli}
& $\checkmark$ 
& $\times$ 
& $\times$ 
& SFT + RL 
& Test stimuli (Tabular) \\

TB or NOT TB~\cite{nadimi2025tb}
& $\checkmark$ 
& $\times$ 
& $\times$ 
& DPO 
& Stimuli Testbench \\

LLM4COV~\cite{zhang2026llm4cov}
& $\checkmark$ 
& $\checkmark$ 
& $\times$ 
& Curriculum SFT 
& Stimuli Testbench \\

\midrule
\textbf{CovR (Ours)} 
& $\checkmark$ 
& $\checkmark$ 
& $\checkmark$ 
& SFT-W/Reason + RL 
& Stimuli Testbench \\
\bottomrule
\end{tabular}
}
\vspace{-2em}
\end{table}

\begin{figure*}[!t]
    \centering
    \includegraphics[width=\textwidth,trim={0cm 2cm 0cm 0cm}]{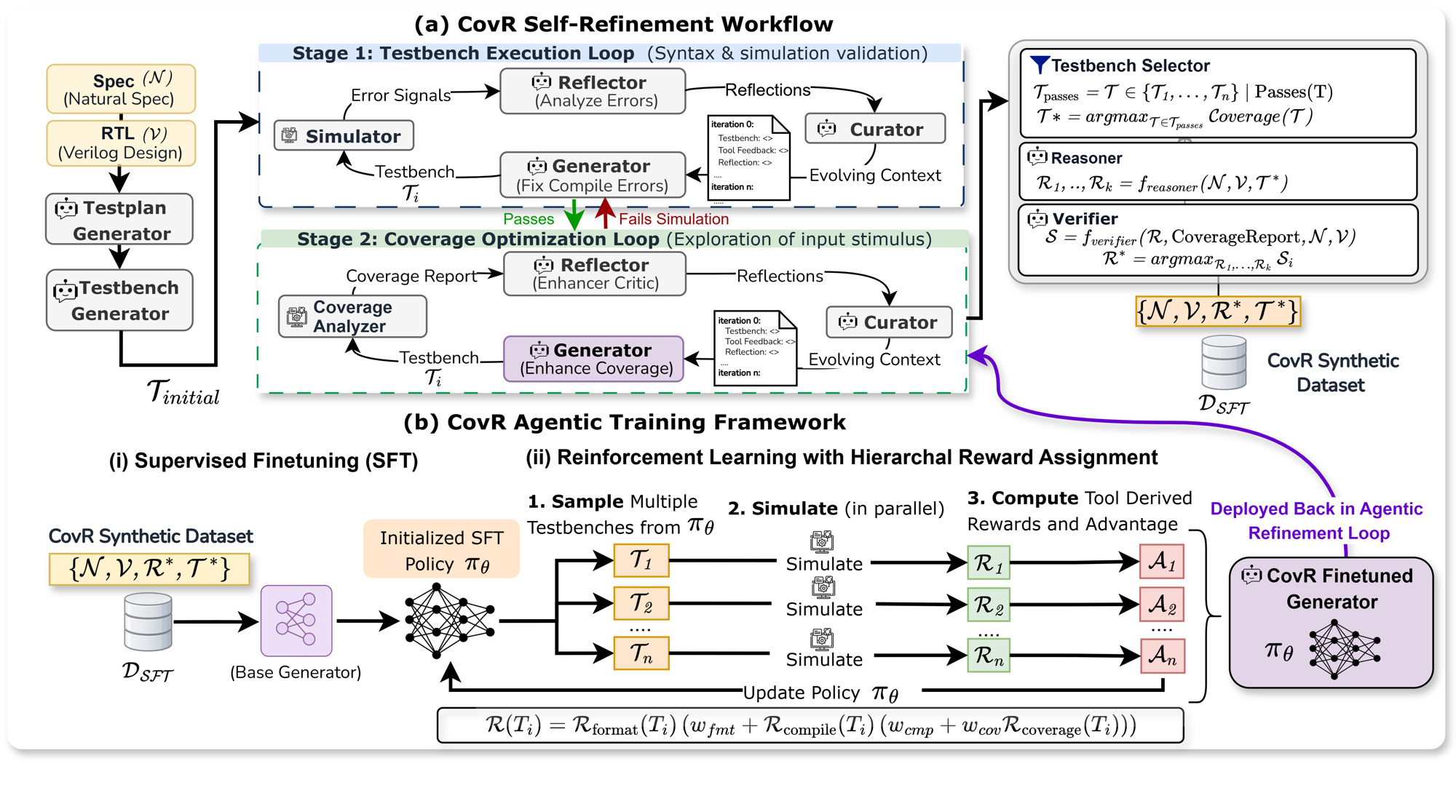}
    \caption{Overview of the \textbf{\emph{CovR}} framework. (a) \emph{Self-refinement workflow:} An initial testbench is generated from the specification $\mathcal{N}$ and RTL $\mathcal{V}$, then iteratively refined through execution and coverage-driven loops to produce a valid high-coverage testbench $\mathcal{T}^*$, which is annotated with reasoning traces $\mathcal{R}^*$. (b) \emph{Agentic training framework:} The resulting tuples $(\mathcal{N}, \mathcal{V}, \mathcal{R}^*, \mathcal{T}^*)$ form a synthetic dataset for supervised fine-tuning (SFT), followed by reinforcement learning (RL), where multiple testbenches are sampled, simulated in parallel, and optimized using hierarchical rewards combining format, compile, and coverage signals.}
    \label{fig:overview}
\end{figure*}

%



Table~\ref{tab:lit_review} summarizes prior approaches. Existing testbench generation frameworks often lack coverage feedback, motivating coverage-aware stimulus generation. Prior coverage-driven methods either rely on limited stimulus representations that hinder scalability to complex designs or lack exploration mechanisms for discovering diverse verification behaviors. \emph{CovR} addresses these limitations through agentic self-refinement and reinforcement learning guided by simulation and coverage feedback.

\section{CovR Framework}
\label{benchmark}

This section presents the \emph{CovR} workflow (Fig.~\ref{fig:overview}), including its agentic pipeline, dataset construction, and learning framework for improving testbench generation. 

\subsection{CovR Self-Refinement Workflow}
Fig.~\ref{fig:overview} (a) presents the \emph{CovR} agentic pipeline. Given a natural language specification $\mathcal{N}$ and RTL design $\mathcal{V}$, the workflow begins with \emph{test plan generation}, which derives a structured testing strategy covering corner cases, reset scenarios, and diverse input conditions. A \emph{Testbench Generator} then produces an initial stimulus $\mathcal{T}_{\text{initial}}$, which is refined through two self-refinement loops: (1) a \emph{testbench execution loop}, which ensures successful testbench simulation, and (2) a \emph{coverage optimization loop}, which uses simulation tool feedback to maximize coverage by targeting uncovered scenarios.

The \emph{testbench execution loop} refines $\mathcal{T}_{\text{initial}}$ to ensure simulation validity. The testbench is compiled and simulated using external tools, and runtime errors are passed to the \emph{Reflector}. The Reflector analyzes the reasoning trace used to generate the testbench and the tool feedback, producing a reflection trace explaining failures in the Generator’s reasoning. This reflection trace, with prior iterations history, is passed to the \emph{Curator}, which maintains an evolving context of failures, fixes, and unresolved issues. This context is then provided to the \emph{Fixer}, which repairs the testbench $\mathcal{T}_i$ using both the latest feedback and prior history. This process repeats until the testbench executes successfully or a maximum number of iterations is reached, after which the workflow transitions to the coverage optimization loop. This iterative improvement process follows the paradigm of \emph{self-refinement} and \emph{self-reflection}~\cite{shinn2023reflexion, zhang2025agentic}, where a model iteratively improves its outputs by leveraging feedback from external tools and its own reasoning traces.

The \emph{coverage optimization loop} improves stimulus quality by targeting hard-to-reach behaviors. Coverage reports from the latest testbench identify unexercised behaviors, including uncovered branches, conditions, and state transitions. Guided by this feedback, the \emph{Generator} refines the testbench by introducing stimuli to close coverage gaps. As in the execution loop, a \emph{Reflector} analyzes coverage deficiencies to guide refinement, while a \emph{Curator} maintains context across iterations. The Curator also emits an \emph{exit} signal to indicate that further refinement is infeasible due to unreachable states. 
If refinement breaks simulation, the workflow returns to the execution loop to restore simulation validity.

Across refinement iterations, multiple testbenches are generated. A \emph{testbench selector} selects the final testbench $\mathcal{T}^*$ by choosing, among all testbenches that are successfully executed, the one achieving the highest coverage. The selected testbench, natural specification $\mathcal{N}$, RTL design $\mathcal{V}$, and coverage report are then passed to the \emph{Reasoner}, which generates candidate reasoning traces describing the verification strategy, stimulus generation, and coverage intent. These traces are evaluated by the  \emph{Verifier} using a structured rubric that measures consistency of the reasoning trace with RTL design, testbench, and the coverage report. The highest-scoring trace is selected as the final explanation $R^*$. The resulting reasoning maps test scenarios to RTL behavior, identifying covered branches, conditions, edge cases, and state transitions, and explaining how coverage is achieved through stimulus. The flow outputs a quadruple of natural language specification $\mathcal{N}$, RTL design $\mathcal{V}$, stimulus testbench $\mathcal{T}^*$, and reasoning trace $\mathcal{R}^*$ explaining how coverage is achieved. The framework supports both open-source (Icarus Verilog~\cite{williams2002iverilog}, Covered~\cite{Covered2010}) and commercial (Synopsys VCS~\cite{synopsys_vcs}, URG~\cite{synopsys_urg}) toolchains.

\vspace{-1em}

\subsection{CovR Synthetic Dataset}

We leverage the \emph{CovR} self-refinement workflow to generate high-coverage testbenches by deploying it with strong foundation models. Specifically, we use \texttt{gpt-4o-mini}~\cite{openai2024gpt4omini} for self-refinement loops and reasoning trace verification, and \texttt{DeepSeek-R1}~\cite{deepseek2025r1} for reasoning trace generation. Each data point is produced via $5$ iterations of self-refinement, with $3$ candidate reasoning traces evaluated per sample. We use Synopsys tools for simulation and coverage feedback to ensure data quality.

\begin{figure}[!t]
    \centering
     \begin{subfigure}{1\linewidth}
      \captionsetup{skip=1pt} 
        \centering
    \includegraphics[width=\linewidth]{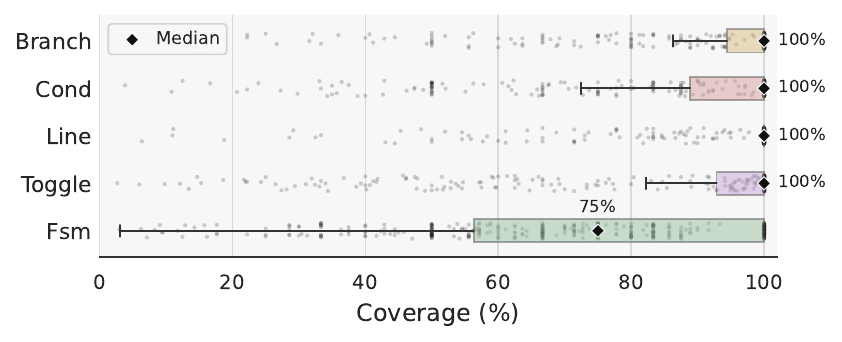}
        \centering
        \label{fig:num_tokens_1}
    \end{subfigure}
    \vspace{-4em}
\caption{\textbf{\emph{CovR}} synthetic dataset coverage distribution. Line, toggle, condition, and branch coverage achieve 100\% median, while FSM coverage is more dispersed with a 75\% median.}
\vspace{-2em}
\label{fig:data_dist}
\end{figure}

Using this pipeline, we construct the \emph{CovR} synthetic dataset from a large collection of RTL–specification pairs sourced from publicly available Verilog code generation datasets. Table~\ref{tab:key_sources} summarizes the key sources, including Pyra~\cite{nadimi2025pyranet} and VeriThoughts~\cite{yubeaton2025verithoughts}, which serve as inputs to the self-refinement workflow. We also incorporate samples from RTLSeek~\cite{zhang2026rtlseek}, VeriPrefer~\cite{wang2025insights}, VeriReason~\cite{wang2025verireason}, and LLM4COV~\cite{zhang2026llm4cov}, converting their testbenches to stimulus-only form and using them as initial seeds for self-refinement to maximize coverage and generate reasoning traces. In total, we generate $16{,}514$ quadruples, comprising a natural language specification, an RTL design, a testbench, and corresponding reasoning trace. Unlike prior coverage-oriented datasets such as LLM4COV~\cite{zhang2026llm4cov}, \emph{CovR} additionally includes reasoning traces that explicitly describe the verification strategy, targeted behaviors, and coverage intent underlying each generated testbench. Fig.~\ref{fig:data_dist} shows the coverage distributions of the \emph{CovR} dataset, showing that most generated testbenches achieve high coverage. Lower-coverage samples are retained in the training set. These capture harder cases where refinement fails or full coverage is inherently infeasible, for example due to constant signals or unreachable states.

 \begin{table}[!t]
\fontsize{7}{6}\selectfont
\centering
\begin{threeparttable}
\captionsetup{skip=2pt}
\caption{RTL and Testbench design sources.}
\setlength{\tabcolsep}{2.8pt}
\renewcommand{\arraystretch}{0.95}

\begin{tabular}{lccccc}
\toprule
Source & Designs & \multicolumn{4}{c}{Statistics \{Median, Max\}} \\
\cmidrule(lr){3-6}
& Count & I/O Ports & \# Cells & Branches & FSM Trans. \\
\midrule

Pyra~\cite{nadimi2025pyranet}
& 1,732 & \{7,3586\} & \{1,16678\} & \{2,274\} & \{9,68\} \\

VeriThoughts~\cite{yubeaton2025verithoughts}
& 8,775 & \{12,21276\} & \{6,3129\} & \{5,527\} & \{6,66\} \\

VeriReason$^{*}$~\cite{wang2025verireason}
& 998 & \{8,256\} & \{3,33046\} & \{3,80\} & \{6,6\} \\

VeriTriplets$^{*}$~\cite{wang2025insights}
& 2,309 & \{54,16772\} & \{36,36488\} & \{16,1151\} & \{7,112\} \\

LLM4Cov$^{*}$~\cite{zhang2026llm4cov}
& 3,468 & \{30,38452\} & \{17,47438\} & \{8,4033\} & \{7,35\} \\

\midrule
\textbf{CovR (ours)}
& \textbf{16,514}
& \textbf{\{19,38452\}}
& \textbf{\{8,47438\}}
& \textbf{\{6,4033\}}
& \textbf{\{6,112\}} \\

\bottomrule
\end{tabular}

\begin{tablenotes}
\scriptsize
\item[*] Refined for coverage and augmented with reasoning traces.
\vspace{-2.8em}
\end{tablenotes}
\label{tab:key_sources}
\end{threeparttable}
\end{table}


\subsection{CovR Training Framework}
The quality of generated testbenches critically depends on the \emph{Generator}'s ability to produce high-coverage stimuli. To enhance this, we propose a two-stage training framework, illustrated in Fig.~\ref{fig:overview} (b), that specializes the \emph{Generator} for coverage-driven testbench generation for later deployment in the agentic framework. First, we perform supervised finetuning (SFT) on the synthetic \emph{CovR} dataset. In this stage, the \emph{Generator} is trained on a dataset $\mathcal{D} = \{(N, V, R^*, T^*)\}$, where each tuple consists of a natural language specification $N$, RTL design $V$, reasoning trace $R^*$, and corresponding testbench $T^*$. Conditioned on $(N, V)$, the model is trained to jointly generate the reasoning trace $R^*$ and the testbench $T^*$ using the standard auto-regressive cross-entropy loss~\cite{radford2019language,bengio2003neural} detailed in Equation~\ref{sft_loss_eq}, where $\mathbf{y}^* = [R^*, T^*]$ is the target output sequence consisting of the reasoning trace followed by the testbench. This stage initializes the model with patterns of high-coverage testbench generation. However, SFT alone may encourage memorization and limit generalization~\cite{chu2025sft}.

%

\begin{table*}[htbp]
\caption{\emph{CovR} versus baseline methods. Evaluation is performed on VerilogEval~\cite{Liu2023verilog}, RTLLM V2.0~\cite{lu2024rtllm}, and CVDP~\cite{pinckney2025comprehensive}.\emph{pass@k} measures simulation validity, and \emph{cov@k} reports the average best coverage. \emph{Agentic} column indicates the inference mode: $\times$ denotes direct model inference without tool feedback, while $\checkmark$ denotes agentic inference with iterative coverage feedback.}
\vspace{-1em}

\centering
\begin{threeparttable}

\resizebox{\textwidth}{!}{%
\begin{tabular}{l|llc|ccc|ccc|ccc|ccc}
\toprule
& \textbf{Model} & \textbf{Variant} & \textbf{Agentic}
& \multicolumn{6}{c|}{\textbf{VerilogEval + RTLLM V2.0}} 
& \multicolumn{6}{c}{\textbf{CVDP (cid012)}} \\
\cmidrule(lr){5-10} \cmidrule(lr){11-16}

& & & 
& \multicolumn{3}{c}{\textbf{pass@k (\%)}} 
& \multicolumn{3}{c|}{\textbf{cov@k (\%)}}
& \multicolumn{3}{c}{\textbf{pass@k (\%)}} 
& \multicolumn{3}{c}{\textbf{cov@k (\%)}} \\

\cmidrule(lr){5-7} \cmidrule(lr){8-10} \cmidrule(lr){11-13} \cmidrule(lr){14-16}

& & &
& \textbf{k=1} & \textbf{k=5} & \textbf{k=10}
& \textbf{k=1} & \textbf{k=5} & \textbf{k=10}
& \textbf{k=1} & \textbf{k=5} & \textbf{k=10}
& \textbf{k=1} & \textbf{k=5} & \textbf{k=10} \\
\midrule

\multirow{3}{*}{Vanilla}

& \texttt{gpt-4o-mini} & -- & $\times$
& 88.62 & 95.25 & 96.06
& 81.27 & 90.01 & 91.60
& 71.08 & 87.15 & 90.30
& 62.10 & 79.78 & 83.75 \\

& \texttt{Qwen2.5-Coder-7B-Instruct} & -- & $\times$
& 77.88 & 93.72 & 95.07
& 67.26 & 86.54 & 89.54
& 51.44 & 75.22 & 80.72
& 41.63 & 65.05 & 71.49 \\

& \texttt{Qwen3-4B-Instruct-2507} & -- & $\times$
& 65.02 & 93.47 & 96.06
& 55.56 & 86.13 & 90.24
& 20.00 & 53.60 & 69.88
& 13.82 & 38.54 & 51.58 \\

\midrule

\multirow{2}{*}{Iterative Feedback}

& \texttt{Qwen2.5-Coder-7B-Instruct} & -- & $\checkmark$
& 80.30 & 90.61 & 92.12
& 73.70 & 84.99 & 87.08
& 69.64 & 82.15 & 84.34
& 56.72 & 70.01 & 73.01 \\

& \texttt{Qwen3-4B-Instruct-2507} & -- & $\checkmark$
& 85.22 & 94.67 & 96.06
& 76.34 & 88.24 & 90.07
& 56.87 & 80.65 & 85.54
& 43.32 & 65.78 & 71.84 \\

\midrule

LLM4COV~\cite{zhang2026llm4cov}

& \texttt{LLM4Cov-Qwen3-4B-SFT-Stage2} & -- & $\times$
& 84.04 & 92.23 & 94.09
& 81.59 & 90.39 & 92.58
& 77.59 & 88.47 & 90.36
& 69.04 & 81.72 & 84.17 \\

\midrule

\multirow{11}{*}{\small \textbf{CovR} (Ours)}

& \texttt{Qwen2.5-Coder-7B-Instruct} & -- & $\checkmark$
& 82.85 & 96.62 & 97.04
& 75.75 & 92.21 & 93.51
& 73.37 & 93.83 & 96.39
& 60.16 & 83.06 & 86.62 \\

& \texttt{Qwen3-4B-Instruct-2507}
& -- & $\checkmark$
& 84.78 & 95.48 & 96.06
& 76.90 & 91.34 & 92.82
& 56.75 & 87.75 & 96.39
& 43.00 & 72.41 & 81.63 \\

\cmidrule(lr){2-16}

& \multirow{4}{*}{\texttt{\textbf{CovR}-Qwen3-4B-Instruct-2507}}

& SFT & $\times$
& 78.91 & 88.41 & 90.15
& 73.53 & 83.56 & 85.65
& 56.75 & 84.62 & 87.95
& 50.78 & 76.98 & 80.85 \\

& & SFT-W/Reason & $\times$
& 77.93 & 93.83 & \underline{96.55}
& 71.34 & 88.49 & 92.07
& 65.42 & 90.98 & \underline{93.98}
& 56.45 & 82.05 & 85.81 \\

& & SFT-W/Reason+RL & $\times$
& 90.79 & \textbf{96.70} & \textbf{97.04}
& 86.50 & 93.19 & 93.81
& \underline{78.92} & \underline{91.41} & \underline{92.77}
& 71.46 & 85.94 & 87.76 \\

& & SFT-W/Reason+RL & $\checkmark$
& \textbf{95.71} & \textbf{97.04} & \textbf{97.04}
& \textbf{91.47} & \underline{94.03} & \underline{94.27}
& \textbf{88.19} & \textbf{93.94} & \textbf{95.18}
& \textbf{79.86} & \textbf{89.66} & \textbf{91.39} \\

\cmidrule(lr){2-16}

& \multirow{4}{*}{\texttt{\textbf{CovR}-Qwen-2.5-coder-7b}}

& SFT & $\times$
& 77.00 & 89.80 & 92.10
& 72.00 & 85.20 & 87.50
& 72.17 & 91.75 & 92.77
& 64.69 & 85.27 & 87.24 \\

& & SFT-W/Reason & $\times$
& 84.50 & \underline{96.51} & \textbf{97.04}
& 78.34 & 91.67 & \underline{92.90}
& 69.04 & 92.24 & 93.98
& 61.12 & 84.21 & 86.79 \\

& & SFT-W/Reason+RL & $\times$
& \underline{92.51} & 96.50 & \underline{96.55}
& \underline{88.49} & \underline{93.29} & \underline{93.61}
& 80.48 & 90.83 & 91.56
& 73.01 & 85.73 & 87.25 \\

& & SFT-W/Reason+RL & $\checkmark$
& \underline{94.38} & \underline{97.03} & \textbf{97.04}
& \underline{90.38} & \textbf{94.10} & \textbf{94.30}
& \underline{87.10} & \underline{94.89} & \underline{95.18}
& \underline{78.72} & \underline{89.40} & \underline{90.44} \\

\bottomrule
\end{tabular}%
}

\begin{tablenotes}
\footnotesize
\item \textbf{Bolded} values indicate the highest value for each column. \underline{Underlined} values indicate the second-highest value. \emph{@k} agentic values are reported by sampling \emph{n=10} \newline different trajectories and refining each trajectory via $5$ iterations.
\end{tablenotes}

\end{threeparttable}
\vspace{-1em}
\label{tab:covr_results}
\end{table*}

\begin{figure*}[!t]
\captionsetup{skip=1pt}
\centering
\captionsetup{font=small}

\begin{subfigure}{0.24\textwidth}
\captionsetup{skip=1pt}
    \centering
    \includegraphics[width=\linewidth]{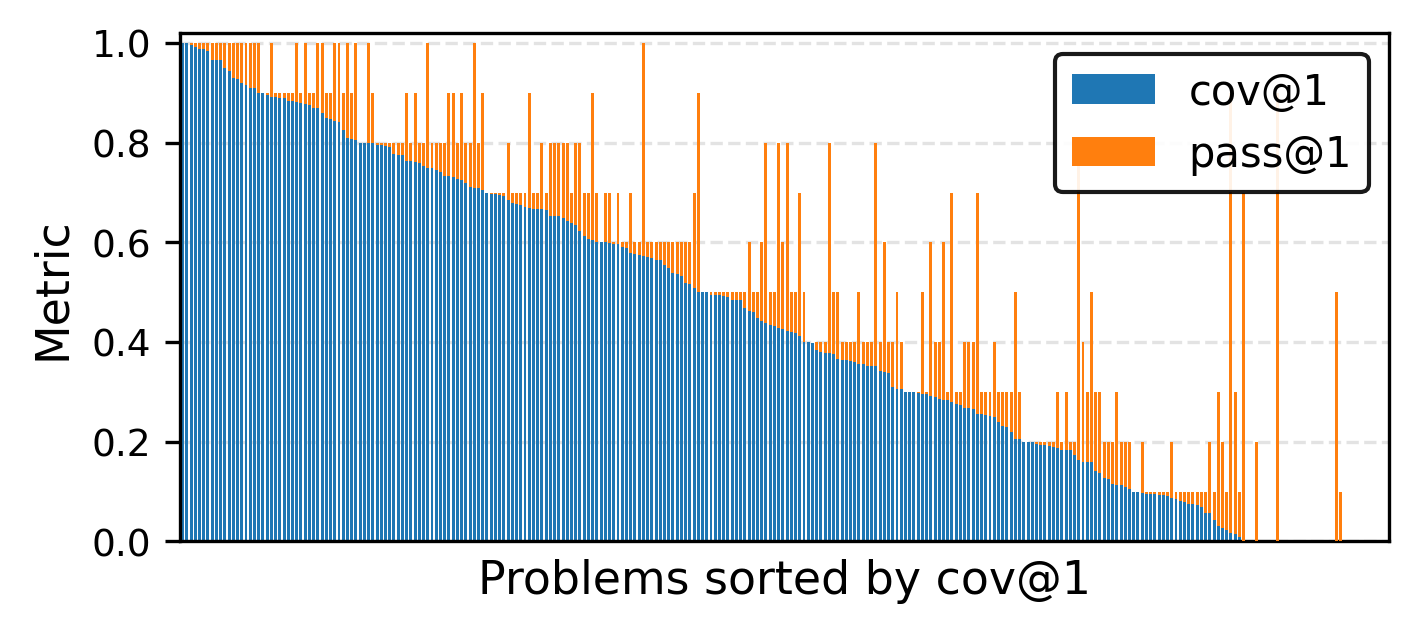}
    \caption{Vanilla}
\end{subfigure}
\hfill
\begin{subfigure}{0.244\textwidth}
\captionsetup{skip=1pt}
    \centering
    \includegraphics[width=\linewidth]{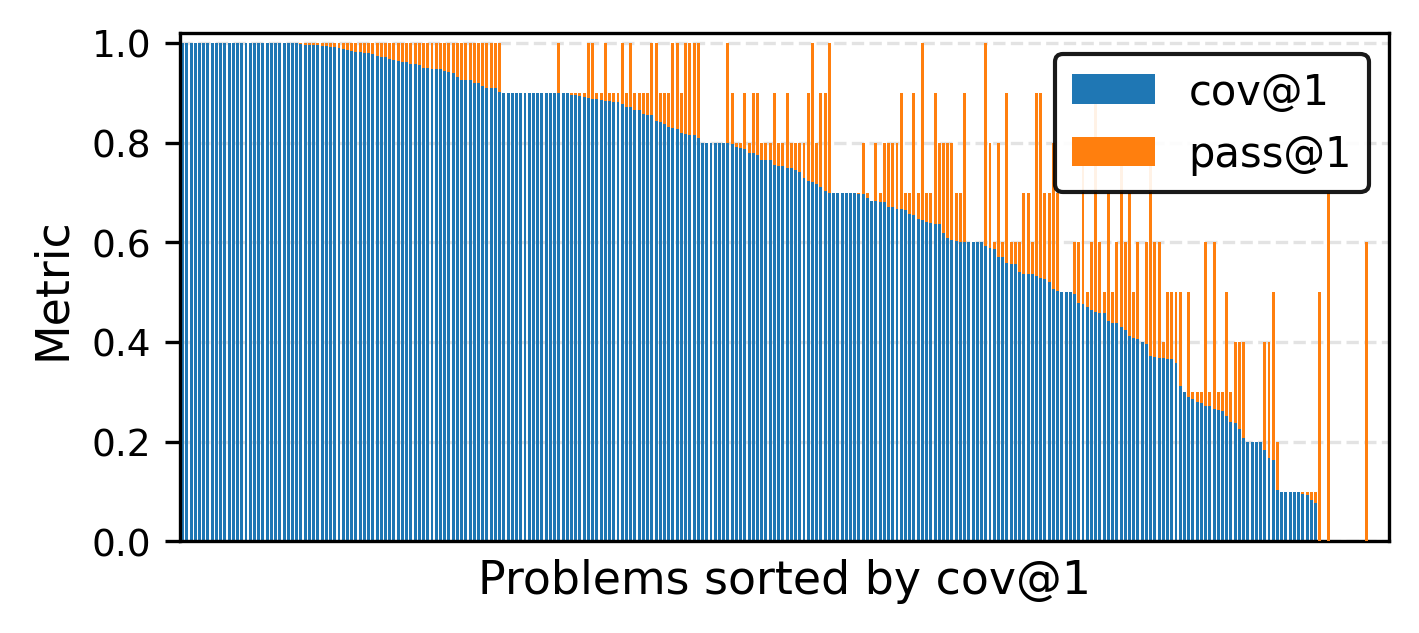}
    \caption{CovR-SFT-W/Reason}
\end{subfigure}
\hfill
\begin{subfigure}{0.24\textwidth}
\captionsetup{skip=1pt}
    \centering
    \includegraphics[width=\linewidth]{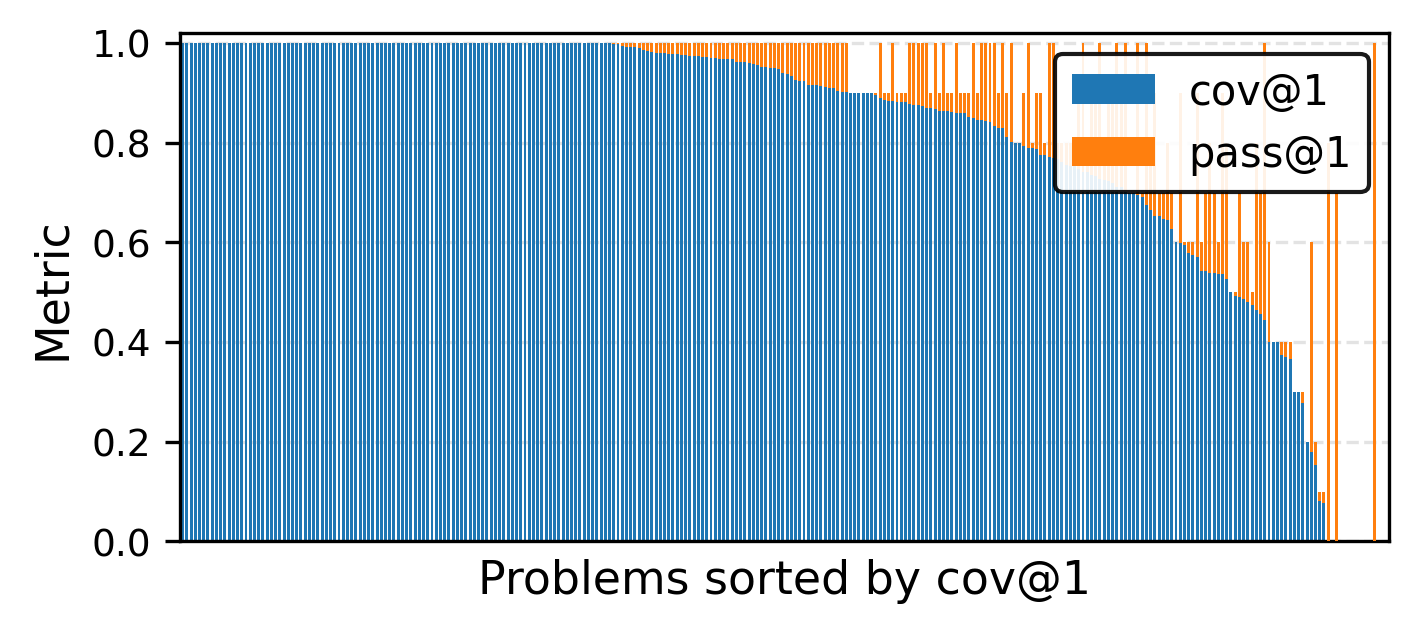}
    \caption{CovR-SFT-W/Reason+RL}
\end{subfigure}
\begin{subfigure}{0.24\textwidth}
\captionsetup{skip=1pt}
    \centering
    \includegraphics[width=\linewidth]{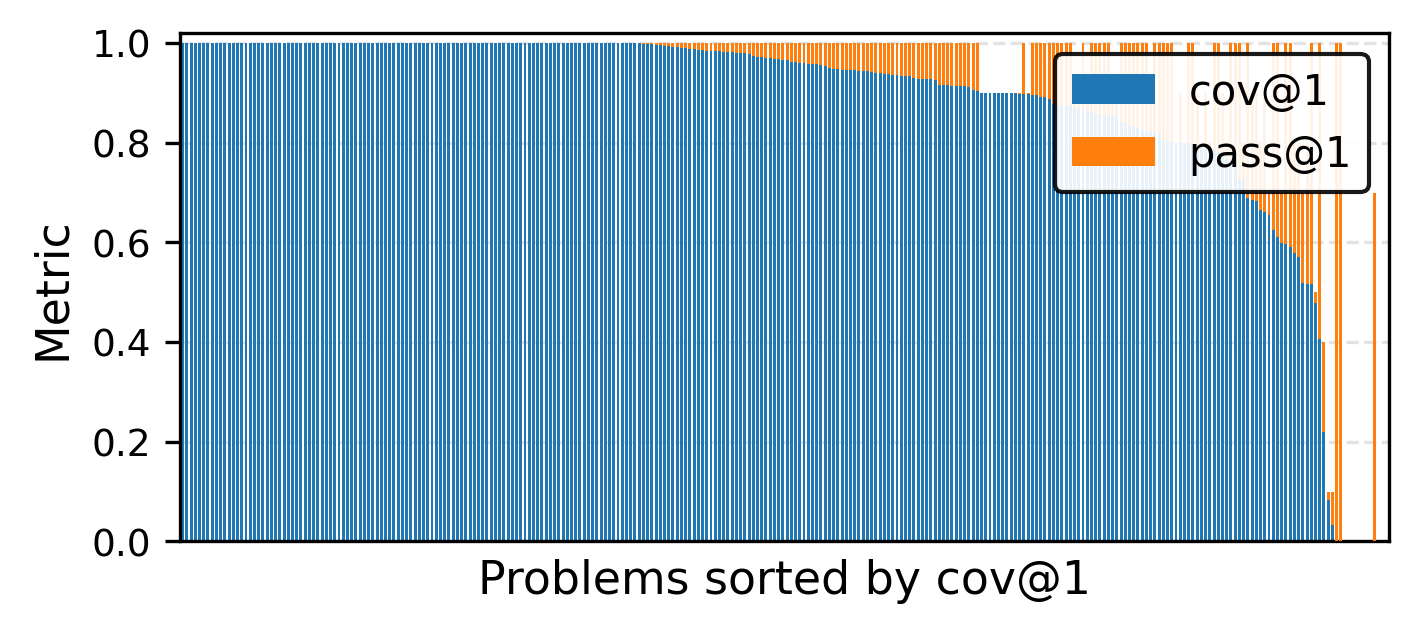}
    \caption{CovR-SFT-W/Reason+RL (Agentic)}
\end{subfigure}
\caption{Distribution of per-problem \emph{pass@$1$} and \emph{cov@$1$} scores for \texttt{Qwen3-4B-Instruct-2507} across (a) the vanilla, (b) reasoning-augmented supervised fine-tuning, (c) GRPO-based reinforcement learning optimized models, and (d) the full agentic pipeline. 
}\label{fig:cov_pass_comparison}
\vspace{-1em}
\end{figure*}

\vspace{-1em}
{\setlength{\abovedisplayskip}{1pt}
 \setlength{\belowdisplayskip}{4pt}
 \setlength{\abovedisplayshortskip}{2pt}
 \setlength{\belowdisplayshortskip}{2pt}
\begin{equation}
\small
\label{sft_loss_eq}
\mathcal{L}_{\text{SFT}}(\theta)
=
-\mathbb{E}_{(N,V,R^*,T^*) \sim \mathcal{D}}
\left[
\sum_{t=1}^{|\mathbf{y}^*|}
\log \pi_{\theta}
\left(
y_t^* \mid N,V,y_{<t}^*
\right)
\right],
\end{equation}
}

To address this limitation, we introduce a reinforcement learning (RL) stage that directly optimizes the \emph{Generator} using tool-derived rewards to maximize testbench coverage. In this stage, the \emph{Generator} is trained with \emph{Group Relative Policy Optimization} (GRPO)~\cite{guo2025deepseek}. Starting from the SFT-initialized policy, GRPO samples a group of candidate testbenches $\{T_1,\dots,T_n\}$ for each input $(N, V)$ and updates the policy based on their relative rewards. Each sampled testbench $T_i$ is executed in a simulator to extract coverage metrics, which are used to compute the rewards. We define a hierarchical reward (Eq.~\ref{reward_eq}) that captures the dependency structure of the verification process. Specifically, $\mathcal{R}_{\text{format}}$ encourages adherence to the expected output structure, $\mathcal{R}_{\text{compile}}$ indicates successful compilation and execution, and $\mathcal{R}_{\text{coverage}}$ measures average code coverage. The weights $w_{\mathrm{cov}} > w_{\mathrm{cmp}} > w_{\mathrm{fmt}}$ prioritize coverage optimization while still encouraging executable testbenches. This formulation ensures that compilation is evaluated only for well-formatted outputs, and coverage is computed only for successfully executed testbenches.

{\fontsize{7.9pt}{8.5pt}\selectfont
\setlength{\abovedisplayskip}{1pt}
 \setlength{\belowdisplayskip}{4pt}
 \setlength{\abovedisplayshortskip}{2pt}
 \setlength{\belowdisplayshortskip}{2pt}
\begin{equation}
\label{reward_eq}
\mathcal{R}(T_i)=
\mathcal{R}_{\mathrm{format}}(T_i)\!\left(
w_{\mathrm{fmt}}
+\mathcal{R}_{\mathrm{compile}}(T_i)\!\left(
w_{\mathrm{cmp}}
+w_{\mathrm{cov}}\mathcal{R}_{\mathrm{coverage}}(T_i)
\right)\right)
\end{equation}
}

Rewards are normalized across the testbenches sampled for the same input $(N,V)$ to compute group-relative advantages (Eq.~\ref{advantage}). Specifically, $\mu_{\mathcal{R}}$ and $\sigma_{\mathcal{R}}$ denote the mean and standard deviation of rewards, and $\delta$ is a small constant to avoid division by zero, $G$ is the number of sampled testbenches. This normalization produces a scale-invariant advantage signal that stabilizes optimization across sampled testbench groups with different reward distributions.


\vspace{-1em}
{\setlength{\abovedisplayskip}{1pt}
 \setlength{\belowdisplayskip}{2pt}
 \setlength{\abovedisplayshortskip}{4pt}
 \setlength{\belowdisplayshortskip}{2pt}
\begin{equation}
\footnotesize
\label{advantage}
\hat{A}_i =
\frac{\mathcal{R}(T_i)-\mu_{\mathcal{R}}}
{\sigma_{\mathcal{R}}+\delta},\;
\mu_{\mathcal{R}}=\frac{1}{G}\sum_{j=1}^{G}\mathcal{R}(T_j),\;
\sigma_{\mathcal{R}}=
\sqrt{\frac{1}{G}\sum_{j=1}^{G}
\left(\mathcal{R}(T_j)-\mu_{\mathcal{R}}\right)^2}.
\end{equation}
}

The final GRPO objective is defined in Eq.~\ref{grpo_loss_eq}. The objective maximizes the likelihood of candidate testbenches that outperform their peers according to the advantage signal $\hat{A}_i$, while constraining policy drift with respect to the supervised finetuned reference policy $\pi_{\mathrm{ref}}$. Here, $\beta$ controls the strength of this regularization, and
$
r_i(\theta)=
\frac{\pi_\theta(T_i \mid N,V)}
{\pi_{\mathrm{ref}}(T_i \mid N,V)}
$
measures how the updated policy changes the likelihood of testbench $T_i$ relative to the reference policy. The clipping operation stabilizes training by limiting large policy updates. As a result, the \emph{Generator} progressively learns to generate simulation-valid testbenches with higher verification coverage.



{\fontsize{6.8pt}{7.5pt}\selectfont
\setlength{\abovedisplayskip}{1pt}
 \setlength{\belowdisplayskip}{4pt}
\begin{equation}
\label{grpo_loss_eq}
\mathcal{L}_{\mathrm{GRPO}}(\theta)=
-\mathbb{E}\!\left[
\frac{1}{G}\sum_{i=1}^{G}
\Big(
\min(
r_i\hat{A}_i,\,
\mathrm{clip}(r_i,1-\epsilon,1+\epsilon)\hat{A}_i
)
-\beta D_{\mathrm{KL}}(\pi_\theta\|\pi_{\mathrm{ref}})
\Big)
\right]
\vspace{-2em}
\end{equation}
}

\section{Experimental Results}
\label{experimental_results}

\subsection{Experimental Setup}
We train \texttt{Qwen-2.5-Coder}~\cite{hui2024qwen25coder} and \texttt{Qwen3-4B-Instruct}~\cite{qwen2025qwen3} on the \emph{CovR} synthetic dataset using our proposed two-stage training framework with LoRA-based adaptation~\cite{hu2022lora} ($r=512$, $\alpha=512$, dropout $=0.1$). All experiments are conducted on NVIDIA B200 GPUs with bfloat16 mixed precision. \textbf{Tools.} For simulation and coverage evaluation, we use Synopsys VCS~\cite{synopsys_vcs} and URG~\cite{synopsys_urg}. \textbf{SFT Stage.} 
We perform supervised finetuning with a learning rate of $2\mathrm{e}{-5}$ and a maximum sequence length of $5\mathrm{K}$ tokens. \textbf{RL Stage.} Starting from the best SFT checkpoint, we further optimize the model using GRPO~\cite{guo2025deepseek}. We use $G=8$ samples per prompt, a maximum sequence length of $4\mathrm{K}$ tokens, a sampling temperature of $1.0$, a constant learning rate of $1\mathrm{e}{-6}$, $\beta=0.01$, and reward weights of $w_{\mathrm{fmt}}=0.02$, $w_{\mathrm{cmp}}=0.13$, and $w_{\mathrm{cov}}=0.85$. \textbf{Evaluation.} All models are evaluated with temperature $0.1$, a maximum sequence length of $32\mathrm{K}$ tokens, and a maximum generation length of $10\mathrm{K}$ tokens. We evaluate model performance on VerilogEval~\cite{Liu2023verilog}, RTLLM V2.0~\cite{lu2024rtllm}, and CVDP (\texttt{cid012})~\cite{pinckney2025comprehensive, zhang2026llm4cov}, comprising 203 tasks for VerilogEval+RTLLM V2.0 and 83 tasks for CVDP.

\subsection{Evaluation Metrics}

\begin{table*}[t]
\captionsetup{skip=2pt}
\caption{CovR as a plug-in stimulus engine for full RTL verification workflows. Integrating CovR improves coverage (cov@1) and mutation detection score (Eval2-100) while preserving simulation validity (Eval0). Lower Eval1 under high coverage stimuli indicates that CovR uncovers previously missed verification bugs.}
\label{tab:covr_integration_split}
\centering
\footnotesize

\setlength{\tabcolsep}{3pt}
\renewcommand{\arraystretch}{0.8}

\resizebox{\textwidth}{!}{
\begin{tabular}{lll|cccc|cccc}
\toprule

& \textbf{Model}
& \textbf{Setting}
& \multicolumn{4}{c|}{\textbf{VerilogEval + RTLLM v2.0}}
& \multicolumn{4}{c}{\textbf{CVDP (cid012)}} \\

\cmidrule(lr){4-7}
\cmidrule(lr){8-11}

&
&
& \textbf{Eval0}
& \textbf{Eval1}
& \textbf{Eval2-100}
& \textbf{cov@1}
& \textbf{Eval0}
& \textbf{Eval1}
& \textbf{Eval2-100}
& \textbf{cov@1} \\

\midrule

\multirow{2}{*}{CorrectBench~\cite{qiu2025correctbench}}


&
\multirow{2}{*}{\texttt{Qwen2.5-Coder-7B}}
& Base
& 43.4
& \textbf{36.0}
& 6.9
& 20.8
& 9.6
& 8.4
& \textbf{1.2}
& 1.4 \\

&
&
\textbf{+CovR (Ours)}
& \textbf{64.5}~\up{21.2}
& 22.2~\textcolor{orange!85!black}{(-13.8)}
& \textbf{9.4}~\up{2.5}
& \textbf{43.1}~\up{22.3}
& \textbf{48.2}~\up{38.6}
& \textbf{14.5}~\up{6.0}
& 0.0~\textcolor{orange!85!black}{(-1.2)}
& \textbf{25.6}~\up{24.2} \\

\midrule

\multirow{2}{*}{PRO-V-R1~\cite{zhao2025pro}}
& \multirow{2}{*}{\texttt{PRO-V-R1-Qwen3-8B}}
& Base
& 93.6
& \textbf{52.2}
& 28.6
& 67.2
& 89.5
& \textbf{42.2}
& 1.2
& 38.2 \\

&
&
\textbf{+CovR (Ours)}
& \textbf{94.1}~\up{0.49}
& 48.3~\textcolor{orange!85!black}{(-3.94)}
& \textbf{32.0}~\up{3.44}
& \textbf{79.3}~\up{12.10}
& \textbf{92.8}~\up{3.30}
& 36.1~\textcolor{orange!85!black}{(-6.10)}
& 1.2~\up{0.00}
& \textbf{55.4}~\up{17.20} \\

\bottomrule
\end{tabular}
}

\vspace{-1em}
\end{table*}

We evaluate generated testbenches along two dimensions: \emph{simulation validity} and \emph{coverage quality}. \textbf{Simulation validity.} We use the standard \emph{pass@k} metric~\cite{chen2021evaluating}, which measures the probability that at least one of the top-$k$ generated testbenches executes successfully without compilation errors or timeouts. \textbf{Coverage quality.} We define a normalized quality score (Eq.~5) that quantifies how closely a generated testbench approaches the maximum attainable coverage $\mathcal{C}^{\max}$. For CVDP, $\mathcal{C}^{\max}$ is defined using the reference coverage targets provided by the benchmark. For problem $i$ and sample $j$, $\hat{\mathcal{C}}_{i,j}$ denotes the reported coverage. The score is represented as a five-dimensional vector (FSM, toggle, line, condition, branch), averaged to a scalar as described below, where $q_{i,j}=1$ indicates full coverage and $q_{i,j}=0$ corresponds to failed simulation or zero coverage.

\vspace{-1em}
\begin{equation}\small
\label{eq:cov}
q_{i,j} =
\begin{cases}
\dfrac{\hat{\mathcal{C}}_{i,j}}{\mathcal{C}^{\max}_i}, & \text{if the sample runs successfully in simulation} \\[6pt]
0, & \text{otherwise.}
\end{cases}
\end{equation}

The quality score is used to compute \emph{cov@k}, defined in Eq.~\ref{eq:covatk}, which measures the expected best coverage achieved when sampling up to $k$ candidate testbenches per problem (i.e., the coverage attainable within $k$ attempts). Here, $N$ is the number of problems, $n$ the number of generated samples per problem, and $J$ a uniformly drawn subset of size $k$. For each problem $i$, $q_{i,j} \in [0,1]$ denotes the quality of sample $j$, and $q_{i,(r)}$ the $r$-th smallest score after sorting the $n$ samples in ascending order. We estimate this expectation using the unbiased subset estimator in Eq.~\ref{eq:covatk}, following prior \emph{eff@k} work~\cite{qiu2024efficient}. The coefficient $\binom{r-1}{k-1}/\binom{n}{k}$ is the probability that the $r$-th ranked sample is the best element in a uniformly selected subset of size $k$. Intuitively, a \emph{cov@k} score of $0.5$ means that the best testbench found within $k$ attempts achieves $50\%$ coverage. The \emph{cov@k} metric is reported per coverage dimension (line, condition, toggle, FSM, branch). Unless otherwise stated, all results in this paper report the scalar \emph{cov@$k$} score obtained by averaging these five dimensions. Due to the cost of simulation tool invocations, all \emph{pass@k} and \emph{cov@k} computations are parallelized across designs and samples.


\vspace{-1em}
\begin{equation}\small
\text{cov@}k = \frac{1}{N}\sum_{i=1}^{N}
\mathbb{E}_{J \subseteq \{1, \ldots, n\},\, |J| = k}\!\left[\max_{j\in J} q_{i,j}\right]
= \frac{1}{N}\sum_{i=1}^{N}\sum_{r=k}^{n}
\frac{\binom{r-1}{k-1}}{\binom{n}{k}} q_{i,(r)}
\label{eq:covatk}
\end{equation}


\subsection{Main Results}

\begin{figure}[!b]
\captionsetup{skip=1pt}
    \centering
\includegraphics[width=\linewidth]{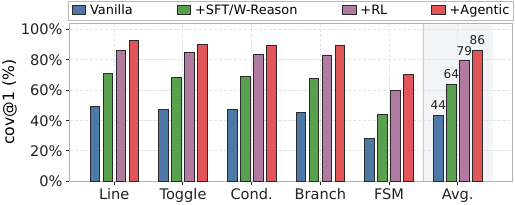}
\centering
\caption{Cov@1 progression for \texttt{Qwen3-4B-Instruct} across inference variants, broken down by coverage metric.}
\vspace{-2em}
    \label{fig:cov_metrics_progression}
\end{figure}


Table~\ref{tab:covr_results} shows that the finetuned \texttt{CovR-Qwen3-4B-Instruct-2507} and \texttt{CovR-Qwen2.5-Coder-7B} models consistently outperform their base counterparts across both benchmarks. Under direct inference, the best-performing \emph{CovR} configuration, \texttt{CovR-Qwen2.5-Coder-7B} with SFT-W/Reason+RL, achieves 88.49\% cov@1 on VerilogEval + RTLLM V2.0 and 73.01\% cov@1 on CVDP, surpassing the teacher model, \texttt{GPT-4o-mini}, by 7.22\% and 10.91\%, respectively. It also outperforms the prior approach, LLM4COV, by 11.85\% and 3.97\% on all benchmarks. When deployed within the agentic self-reflection framework, coverage further improves to 90.38\% cov@1 on VerilogEval + RTLLM V2.0 and 78.72\% cov@1 on CVDP. Compared to the iterative feedback baseline in Table~\ref{tab:covr_results}, which returns tool feedback directly to the \emph{Generator} without self-reflection or context curation, the proposed \emph{CovR} self-reflection loop improves cov@10 by 6.43\% on VerilogEval+RTLLM V2.0 and 13.61\% on CVDP when using the same \texttt{Qwen2.5-Coder-7B} Generator, demonstrating the effectiveness of the Reflector and Curator agents in guiding coverage-oriented refinement. Fig.~\ref{fig:cov_pass_comparison} further illustrates the effectiveness of the proposed training pipeline. As the model progresses from the vanilla baseline through supervised finetuning, reinforcement learning, and agentic deployment, the distributions of \emph{cov@1} and \emph{pass@1} shift consistently toward higher values across all benchmark sets. Fig.~\ref{fig:cov_metrics_progression} further decomposes this progression, highlighting how each stage of the training pipeline contributes to improvements in line, condition, toggle, FSM, and branch coverage.

\subsection{Integration with Verification Workflows}

\emph{CovR} serves as a plug-in stimulus engine for full RTL verification frameworks such as CorrectBench~\cite{qiu2025correctbench} and PRO-V-R1~\cite{zhao2025pro}. These systems target end-to-end verification by combining functional reference models, checking infrastructure, and test-stimuli generation, but they do not explicitly optimize for coverage. In CorrectBench, \emph{CovR} replaces the driver track which generates test stimuli with the stimulus-only testbench generated by \emph{CovR}. In contrast, PRO-V-R1 represents stimuli as a Python-generated \texttt{.json} file specifying DUT input values at each clock cycle. To integrate \emph{CovR} without modifying PRO-V-R1's downstream verification flow, we first simulate the \emph{CovR}-generated testbench, extract the DUT input activity from the resulting \texttt{.vcd} waveform, and convert the recovered input sequence into the \texttt{.json} format expected by PRO-V-R1. Following the evaluation methodology of~\cite{qiu2025correctbench,zhao2025pro}, we report \texttt{Eval0}, which measures successful execution of the generated verification environment; \texttt{Eval1}, which measures the fraction of generated verification environments that pass on the bug-free RTL implementation; and \texttt{Eval2-100}, which measures the percentage of generated verification environments that achieve a perfect mutation score by detecting 100\% of injected RTL mutants. We use MCY-Yosys~\cite{mcy} to generate 10 RTL mutants per design. Results in Table~\ref{tab:covr_integration_split} show that substituting the native Generators with the agentic \texttt{CovR-Qwen2.5-Coder-7B} model consistently improves coverage by 18.95\% and mutation detection score by 1.19\% on average across benchmarks. Interestingly, this reduces \texttt{Eval1} by 4.46\%, indicating that coverage-driven stimuli uncover latent bugs in the generated FRM that were previously undetected.

\section{Conclusion}
\label{conclusion}
In this paper, we present \emph{CovR}, a coverage-aware framework that combines agentic self-refinement with reasoning-guided reinforcement learning to generate high-coverage test stimuli. Experimental results show that the CovR finetuned model achieves 93.81\% cov@10 on VerilogEval and RTLLM V2.0, and 87.76\% cov@10 on CVDP, surpassing GPT-4o-mini by 2.21\% and 4.01\% and outperforming the SOTA baseline LLM4COV by 7.97\% and 3.59\%, respectively. Within full verification workflows, CovR improves achieved coverage by 18.95\%, mutation detection score by 1.19\%, and reveals 4.46\% more previously undetected failures. These results demonstrate the importance of explicitly optimizing coverage in LLM-based hardware verification. Future work includes using asynchronous GRPO~\cite{fu2026areal} to improve scalability, as well as expanding to functional coverage. 

\section*{Acknowledgments}
This work is partially supported by NSF grants 2350180 and 2453413

\bibliographystyle{ACM-Reference-Format}
\bibliography{ref.bib}

\begin{appendix}

\section{Self-Refinement Agents Prompts}
\begin{center}
\begin{tcolorbox}[sharp corners, colback=grey!10, colframe=grey!40!grey, title=Generator (Fix Mode) Prompt, width=\linewidth, boxrule=0.5mm, fontupper=\footnotesize]

You are an expert {language} verification engineer.
Your job is to fix syntax and functional errors and produce a syntactically and functionally valid, simulator-ready testbench.

Non-negotiable requirements:
\begin{enumerate}
    \item {style\_requirements}
    \item Prefer minimal targeted repairs over broad rewrites.
    \item Do not change the intended DUT functionality.
    \item `wire`, `reg`, and `int` declarations are not allowed inside `initial` blocks.
    \item Do not add or keep VCD or waveform dumping logic such as `\$dumpfile` or `\$dumpvars`.
    \item If simulation is timing out, reduce unnecessary stimulus while preserving the intended coverage intent.
\end{enumerate}
Inputs
Fixing Context
\begin{verbatim}
{fixing_context}
\end{verbatim}
 Reflection Summary
\begin{verbatim}
{reflection_summary}
\end{verbatim}

RTL Code
\begin{verbatim}
{rtl_code}
\end{verbatim}

Current Testbench
\begin{verbatim}
{testbench}
\end{verbatim}

Current Compiler/Simulator Error Log
\begin{verbatim}
{testbench_error}
\end{verbatim}

Instructions
Repair the current testbench without weakening its verification quality.
You must preserve or restore the following properties:
\begin{verbatim}
{preservation_requirements}
\end{verbatim}

Use the fixing context as iterative memory:
- avoid repeating failed repair strategies
- preserve helpful prior fixes
- use previous accepted directions when still relevant

Use the reflection summary as high-level guidance.
If the fixing context or reflection summary conflicts with the current compiler/simulator error log, prioritize the current error log.

First briefly reason about:
- the root cause
- which checks must be preserved or repaired
- the minimal safe changes to make

Then output only the final fixed testbench inside a fenced code block exactly like this:
\begin{verbatim}
```testbench
<fixed testbench code here>
```
\end{verbatim}
\end{tcolorbox}
\end{center}

\begin{center}
\begin{tcolorbox}[
    sharp corners,
    colback=grey!10,
    colframe=grey!40!grey,
    title=Generator (Enhance Mode) Prompt,
    width=\linewidth,
    boxrule=0.5mm,
    fontupper=\footnotesize
]
\setlength{\parskip}{1pt}
\setlength{\itemsep}{1pt}

You are an expert Verification Engineer and Testbench Optimizer. Your job is to improve an existing testbench so it closes more coverage on the provided RTL. You are working in an iterative enhancement loop. You may be given memory from previous attempts. Use that memory to avoid repeating ineffective strategies, but always prioritize the current coverage evidence.

\textbf{Important rules:}
\begin{itemize}
    \item[-] Prefer targeted additions over full rewrites.
    \item[-] Focus on uncovered or weakly covered behavior.
    \item[-] Keep the testbench deterministic.
    \item[-] Preserve clocks and resets when valid.
    \item[-] Do not degrade already-covered behavior.
\end{itemize}

\textbf{Enhancement Context}
\begin{verbatim}
{enhancement_context}
\end{verbatim}

\textbf{Reflection Summary}
\begin{verbatim}
{reflection_summary}
\end{verbatim}

\textbf{RTL Code}
\begin{verbatim}
{rtl_code}
\end{verbatim}

\textbf{Current Testbench}
\begin{verbatim}
{testbench}
\end{verbatim}

\textbf{Current Coverage Report}
\begin{verbatim}
{coverage_report}
\end{verbatim}

\textbf{Instructions}

Use enhancement context as memory:
\begin{itemize}
    \item[-] Identify previously attempted directions.
    \item[-] Avoid ineffective changes.
    \item[-] Preserve useful additions.
    \item[-] Reuse successful strategies when relevant.
\end{itemize}

Use reflection summary for:
\begin{itemize}
    \item[-] Remaining coverage gaps.
    \item[-] Promising next scenarios.
    \item[-] Known pitfalls.
\end{itemize}

Prioritize:
\begin{itemize}
    \item[-] Coverage report
    \item[-] RTL + testbench
    \item[-] Reflection summary
    \item[-] Enhancement context
\end{itemize}

If conflicts exist, trust the coverage report.

\textbf{Your task}

Produce an improved testbench that:
\begin{itemize}
    \item[-] Increases coverage with targeted additions.
    \item[-] Preserves existing correct checks.
    \item[-] Avoids unnecessary rewrites.
    \item[-] Adds directed tests, corner cases, assertions, or constrained-random where useful.
    \item[-] Maintains determinism.
    \item[-] Preserves style.
    \item[-] Keeps valid infrastructure (no VCD dumping).
    \item[-] Avoids regression.
\end{itemize}

When useful:
\begin{itemize}
    \item[-] Add concise explanatory comments.
    \item[-] Use fixed-seed constrained random.
    \item[-] Target edge cases and timing-sensitive behavior.
    \item[-] Prefer compact, high-signal additions.
\end{itemize}

\textbf{Output Format}

First, reason about:
\begin{itemize}
    \item[-] Remaining coverage gaps.
    \item[-] What to avoid/reuse.
    \item[-] Planned additions.
\end{itemize}

Then output only:

\begin{verbatim}
```testbench
<enhanced testbench code here>
```

\end{verbatim}
\end{tcolorbox}
\end{center}
\begin{center}
\begin{tcolorbox}[sharp corners, colback=grey!10, colframe=grey!40!grey, title=Reflector (Fix Mode) Prompt, width=\linewidth, boxrule=0.5mm, fontupper=\footnotesize]

Analyze why the testbench failed simulation.

RTL:
\begin{verbatim}
{rtl_code}
\end{verbatim}

Generator Reasoning Trace:
\begin{verbatim}
{reasoning_trace}
\end{verbatim}

Testbench:
\begin{verbatim}
{testbench}
\end{verbatim}

Environment Feedback:
\begin{verbatim}
{testbench_error}
\end{verbatim}

Focus on the concrete compile/simulation blocker. If the issue also hints at a deeper sequencing or DUT-interface misunderstanding, mention it briefly.

**Answer in this exact JSON format:**
\begin{verbatim}
{
  "reasoning": "[Your chain of thought
  / reasoning / thinking process]",
  "error_identification": "[What specifically 
  went wrong in the generator reasoning?]",
  "root_cause_analysis": "[Why did this error 
  occur? What concept was misunderstood?]",
  "correct_approach": "[
  What should the model have done instead?]",
  "key_insight": "[What strategy, formula, 
  or principle should be remembered 
  to avoid this error?]"
}
\end{verbatim}

\end{tcolorbox}
\end{center}
\begin{center}
\begin{tcolorbox}[sharp corners, colback=grey!10, colframe=grey!40!grey, title=Reflector (Enhance Mode) Prompt, width=\linewidth, boxrule=0.5mm, fontupper=\footnotesize]

Analyze why coverage is still insufficient and propose better tests.

RTL:
\begin{verbatim}
{rtl_code}
\end{verbatim}

Generator Reasoning Trace:
\begin{verbatim}
{reasoning_trace}
\end{verbatim}

Testbench:
\begin{verbatim}
{testbench}
\end{verbatim}

Coverage stats:
\begin{verbatim}
{coverage_stats}
\end{verbatim}

Environment Feedback:
\begin{verbatim}
{coverage_report}
\end{verbatim}

When FSM coverage is weak or the RTL appears stateful, explicitly analyze:
\begin{itemize}
    \item[$\bullet$] likely unreached states
    \item[$\bullet$] likely missing transitions
    \item[$\bullet$] guards/prerequisites for those transitions
    \item[$\bullet$] whether additional dwell time or multi-cycle sequencing is required
    \item[$\bullet$] whether the current stimulus is too random, too shallow, or missing directed transition paths
\end{itemize}

Prefer concrete, transition-oriented recommendations such as:
\begin{itemize}
    \item[$\bullet$] reset $\rightarrow$ IDLE $\rightarrow$ START $\rightarrow$ BUSY $\rightarrow$ DONE
    \item[$\bullet$] error/recovery paths
    \item[$\bullet$] timeout paths
    \item[$\bullet$] self-loops with varied dwell length
    \item[$\bullet$] same destination reached from multiple predecessor states
\end{itemize}

**Answer in this exact JSON format:**
\begin{verbatim}
{
  "reasoning": "[Your chain of thought
  / reasoning / thinking process]",
  "error_identification": "[What specifically 
  went wrong in the generator reasoning?]",
  "root_cause_analysis": "[Why did this error 
  occur? What concept was misunderstood?]",
  "correct_approach": "[
  What should the model have done instead?]",
  "key_insight": "[What strategy, formula, 
  or principle should be remembered 
  to avoid this error?]"
}
\end{verbatim}

\end{tcolorbox}
\end{center}
\begin{center}
\begin{tcolorbox}[sharp corners, colback=grey!10, colframe=grey!40!grey, title=Curator Prompt, width=\linewidth, boxrule=0.5mm, fontupper=\footnotesize]

\textbf{Role.} You are an expert iteration-memory curator for an agentic RTL verification pipeline. Your sole responsibility is to maintain a concise and structured evolving context across iterations. You do not generate RTL, testbenches, or reflections, and you do not provide free-form advice.

\textbf{Inputs.} You receive the existing evolving context, the latest generator output summary, tool feedback, reflection summary, and acceptance signal.

\textbf{Purpose.} Maintain cross-iteration memory in a compact, structured form, capturing what was attempted, what failed or improved, reflection insights, acceptance status, and whether progress has stalled.

\textbf{Rules.}
\begin{itemize}
    \item Preserve all prior iteration records from the evolving context.
    \item Append exactly one new iteration record.
    \item Do not delete prior iterations unless the context is empty.
    \item Summarize concisely; avoid copying raw logs.
    \item Do not invent outcomes; summarize conservatively.
    \item Mark \texttt{stalled} if recent iterations show no meaningful progress.
    \item Set \texttt{restart\_recommended} only if the trajectory is clearly stuck.
    \item Set \texttt{exit\_recommended} only if further iterations are unlikely to help.
\end{itemize}

\textbf{Output Schema.} Return valid JSON with exactly the following structure:
\begin{verbatim}
{
  "iterations": [
    {
      "iteration": 1,
      "mode": "fix",
      "generator_output": "short summary",
      "errors_encountered": "short summary",
      "llm_reflection": "short summary",
      "accepted": false
    }
  ],
  "current_status": {
    "latest_issue": "short summary",
    "stalled": false,
    "restart_recommended": false,
    "exit_recommended": false
  }
}
\end{verbatim}

\textbf{Guidelines.}
\begin{itemize}
    \item Keep summaries concise and factual.
    \item \texttt{generator\_output}: describe changes, not full code.
    \item \texttt{errors\_encountered}: summarize key failures or coverage gaps.
    \item \texttt{llm\_reflection}: 1--3 sentence distilled insight.
    \item \texttt{accepted}: whether the iteration improved sufficiently.
    \item \texttt{latest\_issue}: most critical unresolved problem.
\end{itemize}

\textbf{Existing Context:}
\begin{verbatim}
{evolving_context}
\end{verbatim}
\textbf{Iteration:} \texttt{\{iteration\}}

\textbf{Generator Output:}
\begin{verbatim}
{generator_output}
\end{verbatim}
\textbf{Tool Feedback:}
\begin{verbatim}
{tool_feedback}
\end{verbatim}
\textbf{Reflection Summary:}
\begin{verbatim}
{reflection_summary}
\end{verbatim}
\textbf{Accepted:} \texttt{\{accepted\}}

\textbf{Instructions.}
\begin{enumerate}
    \item Preserve all prior iterations.
    \item Append one new iteration record.
    \item Summarize the current output, issues, and reflection concisely.
    \item Update \texttt{current\_status}.
    \item Detect stagnation and update flags accordingly.
    \item Return valid JSON only. Do not include any additional text.
\end{enumerate}

\end{tcolorbox}
\end{center}
\begin{center}
\begin{tcolorbox}[sharp corners, colback=grey!10, colframe=grey!40!grey, title=Reasoner Prompt, width=\linewidth, boxrule=0.5mm, fontupper=\footnotesize]

Your task is to INFER the internal reasoning process of a verification engineer.

Given the RTL design, the final testbench, reconstruct the step-by-step verification reasoning that would naturally lead to writing this testbench.

\textbf{IMPORTANT:}
\begin{itemize}
    \item The reasoning must be written as if it precedes the testbench.
    \item Do not describe the testbench at a surface level. Instead, explain the intent behind each verification decision: why specific inputs are chosen, why certain behaviors are tested, how they relate to the RTL logic, and how they help exercise coverage-relevant behaviors.
    \item Every stimulus described must be explicitly linked to a concrete RTL condition, such as an \texttt{if}/\texttt{else} branch, case item, reset path, state transition, corner case, or output condition.
    \item Do not introduce behaviors, checks, branches, states, or coverage claims that are not supported by the RTL, testbench, or coverage evidence.
    \item Emphasize why the selected stimuli are useful for covering distinct RTL behaviors, not just for functional validation.
    \item If important RTL behaviors remain uncovered, briefly mention them as limitations of the first-pass reasoning, but do not invent tests that are not present.
    \item Do not narrate the testbench (e.g., ``the testbench applies \ldots''). Express each point as a verification decision rather than a description of actions.
    \item Coverage discussion must refer to distinct RTL behaviors rather than generic coverage terms.
\end{itemize}

This is NOT an explanation task.
This is a reconstruction of the reasoning process that leads to the testbench.

The reasoning should reflect a coverage-aware verification mindset:
identify how a verification engineer would choose inputs and checks to exercise meaningful and distinct RTL behaviors, not just to observe outputs.

Use concise, technical, verification-oriented language.
Prefer short numbered steps or compact paragraphs.

For each major test stimulus, explain both:
\begin{enumerate}
    \item the functional intent, and
    \item the distinct RTL behavior or coverage target it is meant to exercise.
\end{enumerate}

Use a concise structured format.

Focus on the most important reasoning only. Do not add routine DUT-instantiation discussion unless it is unusual or materially relevant to the verification intent.

Structure each reasoning trace using only the relevant sections below:

\begin{enumerate}
    \item \textbf{Verification Goal}  
    What functionality or behavior needs to be validated.

    \item \textbf{Stimulus Design}  
    Why the important inputs, sequences, or scenarios are chosen and which RTL behaviors they are intended to exercise. Highlight when a stimulus targets something specific such as a branch, case split, reset path, FSM transition, or boundary condition.

    \item \textbf{Timing and Execution}  
    Include only timing, reset, or sampling details that materially affect behavior.

    \item \textbf{Coverage Intent and Corner Cases}  
    Identify the distinct RTL behaviors and edge cases being targeted, and briefly note important uncovered behaviors if present.
\end{enumerate}

Keep each section concise and grounded in the RTL and testbench.
Omit any section that is not relevant instead of adding generic content.
"""

Natural language specification:
\begin{verbatim}
    {natural_spec}
\end{verbatim}

RTL code:
\begin{verbatim}
```verilog
{verilog}
```  
\end{verbatim}

Testbench code:
\begin{verbatim}
```verilog
{testbench}
```
\end{verbatim}

\end{tcolorbox}
\end{center}

\begin{center}
\begin{tcolorbox}[sharp corners, colback=grey!10, colframe=grey!40!grey, title=Verifier Prompt, width=\linewidth, boxrule=0.5mm, fontupper=\footnotesize]

You are an expert verification reviewer acting as an LLM judge.

You will score each candidate reasoning trace based on how well it explains the provided testbench with respect to the natural language specification, RTL, and coverage report.

Return exactly a JSON object with the following schema:
\begin{verbatim}
{{
  "ranking": [
    {{
      "id": "trace_1",
      "score": 9,
      "reason": "justification grounded
      in the testbench behavior 
      and coverage report"
    }}
  ]
}}

\end{verbatim}

Judging criteria:
\begin{itemize}
    \item correctness with respect to the RTL and testbench behavior
    \item specificity and grounding in actual testbench logic
    \item effective use of coverage report evidence
    \item usefulness for testbench-generation reasoning data
    \item clarity, conciseness, and non-redundancy
    \item penalize hallucinations or unsupported claims
\end{itemize}

Ground all judgments in the RTL, testbench, and coverage report. Prefer traces that reference concrete signals, conditions, or behaviors. Do not reward vague or generic explanations.

Do not return markdown fences or additional text.

Natural language specification:
\begin{verbatim}
{natural_spec}
\end{verbatim}

RTL code:
\begin{verbatim}
{verilog}
\end{verbatim}

Testbench code:

\begin{verbatim}
{testbench}
\end{verbatim}

Coverage report:
\begin{verbatim}
{coverage_report}
\end{verbatim}

Candidate reasoning traces:
\begin{verbatim}
{reasoning_traces}
\end{verbatim}

\end{tcolorbox}
\end{center}

\section{CovR Synthetic Data}
Figure ~\ref{fig:sample} shows a sample from the \emph{CovR} synthetic data constructed from the self-refinement workflow. 
\begin{figure}[H]
    \centering
    \begin{subfigure}{\linewidth}
        \centering

\begin{tcolorbox}[covrbox, title=Natural Specification ($\mathcal{N}$), before skip=0pt]
    \setlength{\parindent}{0pt}
    \setlength{\parskip}{0pt}
\begin{verbatim}
Span4Mux_h4 module has the following interface: input I (4-bit vector); 
input S (2-bit vector); 
output O (single bit). 
The module should function as a 4-to-1 multiplexer. The input S selects 
one of the bits from the input I to be routed to the output O... 
\end{verbatim}
\end{tcolorbox}

\begin{tcolorbox}[covrbox, title=RTL Design ($\mathcal{V}$)]
    \begingroup
    \setlength{\parindent}{0pt}
    \setlength{\parskip}{0.6pt}
\begin{lstlisting}[style=VerilogStyle]
module Span4Mux_h4(input [3:0] I, input [1:0] S, output O);
  assign O = (S == 2'b00) ? I[0] : 
             (S == 2'b01) ? I[1] :
             (S == 2'b10) ? I[2] : I[3];
endmodule
\end{lstlisting}
    \par
    \endgroup
\end{tcolorbox}

\begin{tcolorbox}[covrbox, title=Reasoning Trace ($\mathcal{R^*}$)]
    \begingroup
    \setlength{\parindent}{0pt}
    \setlength{\parskip}{0.6pt}

\textbf{Verification Goal}  
Validate that for every possible 2-bit select value `S`, the output `O` corresponds exactly to the selected bit of the 4-bit input `I`, matching the conditional assignment in the RTL.

\textbf{Stimulus Design} 1. \textbf{All-zero and all-one I patterns} – Apply `I=4'b0000` and `I=4'b1111` for each `S` value (00, 01, 10, 11).  
   - **Functional intent:** Verify the multiplexer’s basic routing for both logic-0 and logic-1 across all selection branches.  
   - **RTL target:** Exercise each of the four conditional branches (`S == 2’b00`, `2’b01`, `2’b10`, `2’b11`) with inputs that cause the selected bit to be 0 and 1, ensuring each branch propagates the correct logic level.  2. \textbf{Mixed I patterns with specific bit emphasis} – Apply `I=4'b1000` (only MSB=1) with `S=00`, `I=4'b0111` (only LSB=0) with `S=01`, `I=4'b0001` (only LSB=1) with `S=10`, `I=4'b1110` (only LSB=0) with `S=11`.  
   - **Functional intent:** Verify that each selection path correctly isolates the intended bit when other bits differ, catching potential cross-coupling or bit-selection errors.  
   - **RTL target:** For each branch, ensure the condition picks the exact indexed bit (`I[0]`, `I[1]`, `I[2]`, `I[3]`), not an adjacent or constant value.  

3. \textbf{Random I and S combinations} – Generate 10 random `I` and `S` values.  
   - **Functional intent:** Stress the mux with arbitrary input combinations beyond directed patterns.  
   - **RTL target:** Cover a broader subset of the 64 possible `{I,S}` combinations, increasing confidence that no unexpected interaction exists between `I` bits and the select logic.  
4. \textbf{Rapid S changes with constant I} – Hold `I=4'b1010` and sequence `S` through 00→01→10→11 with short delays.  
   - **Functional intent:** Verify the combinational path responds immediately to select changes, and that the output correctly toggles between 0 and 1 as different bits are selected.  
   - **RTL target:** Exercise the conditional assignment’s switching behavior as `S` changes, ensuring no lingering or transient output values.

\textbf{Timing and Execution}  
- Primary test intervals use 100-time-unit delays, allowing clear observation of steady-state outputs.  
- Rapid-change sequence uses 10-time-unit delays to check dynamic response without requiring timing-model consideration (purely combinational).  

\textbf{Coverage Intent and Corner Cases}  
- \textbf{Covered} All four select branches with both 0 and 1 on the selected bit; mixed-bit patterns to ensure bit-indexing correctness; random combinations for broader input-space sampling; quick switching to verify combinational behavior.  
- **Limitations:** No explicit test for unknown/high-impedance values on `I` or `S`; no check for glitches during simultaneous `I` and `S` changes; only 10 random samples leave many of the 64 possible combinations uncovered. Directed tests focus on functional correctness of each branch but do not systematically cover all `I` bit-value combinations per branch.

    \endgroup
\end{tcolorbox}

\begin{tcolorbox}[covrbox, title=Testbench ($\mathcal{T^*}$)]
    \begingroup
    \setlength{\parindent}{0pt}
    \setlength{\parskip}{0.6pt}
\begin{lstlisting}[style=VerilogStyle]
module testbench;
  reg [3:0] I;
  reg [1:0] S;
  wire O;
  
  Span4Mux_h4 uut (.I(I), .S(S), .O(O));
  
  initial begin
    // Directed tests for specific edge cases
    I = 4'b0000; S = 2'b00; #100; 
    I = 4'b0000; S = 2'b01; #100; 
    I = 4'b0000; S = 2'b10; #100; 
    I = 4'b0000; S = 2'b11; #100; 
    
    I = 4'b1111; S = 2'b00; #100; 
    I = 4'b1111; S = 2'b01; #100; 
    I = 4'b1111; S = 2'b10; #100; 
    I = 4'b1111; S = 2'b11; #100; 
    
    I = 4'b1000; S = 2'b00; #100; 
    I = 4'b0111; S = 2'b01; #100; 
    I = 4'b0001; S = 2'b10; #100; 
    I = 4'b1110; S = 2'b11; #100; 
    
    // Random tests
    repeat (10) begin
      I = $random % 16; // Random I value from 0 to 15
      S = $random % 4;  // Random S value from 0 to 3
      #100; 
    end
    
    // Rapidly change S while keeping I constant
    I = 4'b1010; // Set a constant I
    S = 2'b00; #10; 
    S = 2'b01; #10; 
    S = 2'b10; #10; 
    S = 2'b11; #10; 
    
    // Timeout
    #1000 $finish;
  end
endmodule
\end{lstlisting}

{\fontsize{6}{6}\selectfont\textit{Coverage Achieved: 100\% cond, 100\% toggle, 100\% branch}}
    \endgroup
\end{tcolorbox}

    \end{subfigure}    

\caption{\emph{CovR} dataset sample, showing high coverage testbench and its corresponding reasoning annotations.}
\label{fig:sample}  
\end{figure}

\section{Efficient Pass@k and Cov@k Computation}

The \emph{pass@k} metric is defined in Eq.~\ref{eq:pass@k}, where $N$ is the number of designs, $k$ is the number of allowed attempts, $n_i$ is the number of generated samples for problem $i$, and $c_i$ is the number of samples that execute successfully in simulation. Intuitively, \emph{pass@k} measures the fraction of problems for which at least one of the top-$k$ generated testbenches runs successfully.

\begin{equation}
\label{eq:pass@k}
\text{pass@}k
= \mathbb{E}_{i=1}^{N}\!\left[\,1-\frac{C\!\left(n_i-c_i,\,k\right)}{C\!\left(n_i,\,k\right)}\right]
\end{equation}

The \emph{cov@k} is defined in Eq.~\ref{eq:covatk}. Both \emph{pass@k} and \emph{cov@k} require sampling multiple testbenches per design and executing each sample with EDA tools for simulation and coverage extraction. As these tool invocations are computationally expensive, we design the \emph{CovR} evaluation system for efficient, large-scale parallel execution.

\begin{figure}[!t]
\centering
\resizebox{\columnwidth}{!}{%
\begin{tikzpicture}[
    font=\sffamily,
    >=Latex,
    node distance=0.7cm,
    box/.style={
        draw,
        rounded corners=3pt,
        thick,
        align=center,
        minimum width=1.6cm,
        minimum height=0.9cm
    },
    smallbox/.style={
        draw,
        rounded corners=2pt,
        align=center,
        minimum width=1.25cm,
        minimum height=0.28cm,
        font=\tiny
    },
    gpu/.style={box, draw=blue!70!black, text=blue!70!black},
    cpu/.style={box, draw=green!50!black, text=green!50!black},
    metric/.style={box, draw=purple!70!black, text=purple!70!black},
    db/.style={
        cylinder,
        draw,
        thick,
        shape border rotate=90,
        aspect=0.28,
        minimum height=1.5cm,
        minimum width=1.1cm,
        align=center
    },
    arrow/.style={->, thick}
]

\node[font=\bfseries\small, text=blue!70!black] (s1) at (-2.7,3.0)
{\textcircled{\scriptsize 1} LLM Inference};

\node[font=\bfseries\small, text=green!50!black] (s2) at (0.9,3.0)
{\textcircled{\scriptsize 2} Coverage};

\node[font=\bfseries\small, text=purple!70!black] (s3) at (4.3,3.0)
{\textcircled{\scriptsize 3} Metrics};

\draw[arrow] (-1.2,3.0) -- (-0.2,3.0);
\draw[arrow] (2.3,3.0) -- (3.2,3.0);

\node[draw, rounded corners=3pt, minimum width=1.5cm,
minimum height=3.6cm] (problems) at (-4.2,0) {};

\node[font=\bfseries\scriptsize] at (-4.2,1.45) {Problems};

\node[smallbox] at (-4.2,0.95) {problem 1};
\node[smallbox] at (-4.2,0.45) {problem 2};
\node[font=\small] at (-4.2,-0.05) {$\vdots$};
\node[smallbox] at (-4.2,-0.55) {problem b};
\node[smallbox] at (-4.2,-1.05) {problem bz};

\node[gpu] (gpu0) at (-1.9,0.9)
{\bfseries GPU 0\\[-1pt]\scriptsize Batch};

\node[gpu] (gpun) at (-1.9,-0.9)
{\bfseries GPU n\\[-1pt]\scriptsize Batch};

\node[font=\Large] at (-1.9,0) {$\vdots$};

\node[db] (db) at (0.4,0)
{\bfseries Answers\\[-1pt]\scriptsize JSONL};

\node[cpu] (cpu0) at (2.8,0.9)
{\bfseries CPU 0\\[-1pt]\scriptsize VCS};

\node[cpu] (cpun) at (2.8,-0.9)
{\bfseries CPU m\\[-1pt]\scriptsize VCS};

\node[font=\Large] at (2.8,0) {$\vdots$};

\node[metric, minimum width=1.7cm, minimum height=2.2cm]
(metrics) at (5.2,0)
{
\bfseries\scriptsize Metrics\\[4pt]
\fbox{\scriptsize pass@k}\\[5pt]
\fbox{\scriptsize cov@k}
};

\draw[arrow] (problems.east |- gpu0.west) -- (gpu0.west);
\draw[arrow] (problems.east |- gpun.west) -- (gpun.west);

\draw[thick] (gpu0.east) -- ++(0.35,0) |- (db.west);
\draw[thick] (gpun.east) -- ++(0.35,0) |- (db.west);

\draw[thick] (db.east) -- ++(0.35,0) |- (cpu0.west);
\draw[thick] (db.east) -- ++(0.35,0) |- (cpun.west);

\draw[thick] (cpu0.east) -- ++(0.35,0) |- (metrics.west);
\draw[thick] (cpun.east) -- ++(0.35,0) |- (metrics.west);

\end{tikzpicture}
}
\vspace{-0.5em}
\caption{Efficient evaluation pipeline of \emph{CovR}. Problems are distributed across GPUs for LLM inference, stored in JSONL format, and evaluated in parallel using Synopsys VCS to compute pass@k and cov@k metrics.}
\label{fig:covr_eval}
\vspace{-1em}
\end{figure}
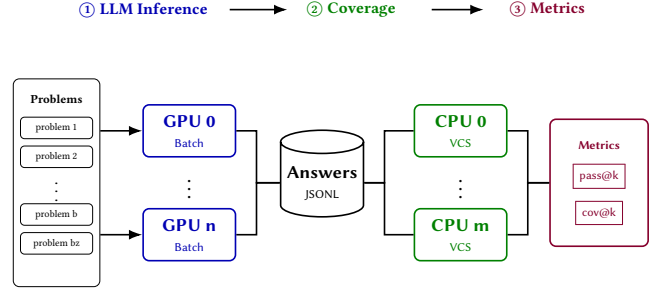

Since EDA tool calls are computationally expensive, we design the \emph{CovR} evaluation system to support parallel execution for efficient \emph{pass@k} and \emph{cov@k} computation. To address this, we develop a fast, highly parallelized evaluation pipeline that decomposes the workflow into two stages: testbench generation and evaluation, as illustrated in Fig ~\ref{fig:covr_eval}.

\section{CovR Kills Mutations and Reveals Latent Verification Failures}

Figure~\ref{fig:covr-mutant-killing-scancode} shows how CovR’s broader stimulus kills more mutants than PRO-V~\cite{zhao2025pro} stimulus engine.


\begin{figure}[H]
\centering
\scriptsize

\definecolor{covrgreen}{RGB}{34,120,65}
\definecolor{provred}{RGB}{150,50,45}
\definecolor{panelgray}{RGB}{248,248,248}
\definecolor{mutblue}{RGB}{34,76,120}
\definecolor{misshighlight}{RGB}{238,238,238}

\newtcbox{\mutantid}{
  on line,
  colback=black,
  colframe=black,
  coltext=white,
  boxrule=0pt,
  arc=0.8mm,
  left=2pt,
  right=2pt,
  top=1pt,
  bottom=1pt,
  boxsep=0pt,
  fontupper=\bfseries\tiny\ttfamily
}

\newcommand{\killchip}[1]{%
  \tcbox[
    on line,
    colback=covrgreen!12,
    colframe=covrgreen!75!black,
    coltext=covrgreen!45!black,
    boxrule=0.2pt,
    arc=0.4mm,
    left=1.2pt,right=1.2pt,top=0.2pt,bottom=0.2pt
  ]{\scriptsize\ttfamily #1}%
}

\newcommand{\survivechip}[1]{%
  \tcbox[
    on line,
    colback=provred!12,
    colframe=provred!80!black,
    coltext=provred!60!black,
    boxrule=0.2pt,
    arc=0.4mm,
    left=1.2pt,right=1.2pt,top=0.2pt,bottom=0.2pt
  ]{\scriptsize\ttfamily #1}%
}

\newcommand{\mutheading}[1]{%
\vspace{1pt}
{\color{black!20}\rule{\linewidth}{0.25pt}}
\vspace{-2pt}

{\bfseries\tiny\color{black!70} #1}
\vspace{-2pt}
}

\newcommand{\mutline}[2]{%
{\ttfamily\tiny #1\hspace{2pt}#2}\par\vspace{1.5pt}
}

\newcommand{\mutlinehl}[2]{%
\noindent\colorbox{misshighlight}{%
  \parbox{\dimexpr\linewidth-2\fboxsep\relax}{%
    \ttfamily\tiny #1\hspace{2pt}#2%
  }%
}\par\vspace{1.5pt}
}

\lstdefinestyle{verilogstyle}{
  language=Verilog,
  basicstyle=\ttfamily\tiny,
  keywordstyle=\color{blue!70!black}\bfseries,
  commentstyle=\color{green!40!black},
  stringstyle=\color{orange!70!black},
  breaklines=true,
  columns=fullflexible,
  keepspaces=true,
  showstringspaces=false,
  frame=none,
  xleftmargin=0.25em,
  aboveskip=1pt,
  belowskip=1pt
}

\tcbset{
  enhanced,
  boxsep=1pt,
  left=2pt,
  right=2pt,
  top=2pt,
  bottom=2pt,
  arc=0.6mm,
  boxrule=0.35pt,
  colframe=black!60,
  colback=white,
  fonttitle=\bfseries\scriptsize,
  before skip=2pt,
  after skip=2pt
}

\begin{tcolorbox}[title={(a) Specification}, colback=panelgray]
\texttt{TopModule} decodes four keyboard scancodes into one-hot direction
signals:
\[
\texttt{E06B}\!\rightarrow\!\texttt{left},\quad
\texttt{E072}\!\rightarrow\!\texttt{down},\quad
\texttt{E074}\!\rightarrow\!\texttt{right},\quad
\texttt{E075}\!\rightarrow\!\texttt{up}.
\]
All other scancodes should produce
\texttt{left=down=right=up=0}.
\end{tcolorbox}

\vspace{-4pt}

\begin{minipage}[t]{0.495\columnwidth}
\begin{tcolorbox}[title={(b) Full Benchmark RTL}, equal height group=A]
\begin{lstlisting}[style=verilogstyle]
module TopModule (
  input [15:0] scancode,
  output reg left,
  output reg down,
  output reg right,
  output reg up
);

always @(*) begin
  {up, left, down, right} = 0;

  case (scancode)
    16'he06b: left  = 1;
    16'he072: down  = 1;
    16'he074: right = 1;
    16'he075: up    = 1;
  endcase
end

endmodule
\end{lstlisting}
\end{tcolorbox}
\end{minipage}
\hfill
\begin{minipage}[t]{0.495\columnwidth}
\begin{tcolorbox}[
  title={(c) Representative Mutant Faults},
  equal height group=A
]



\mutheading{Inverted decode predicates}

\begin{lstlisting}[style=verilogstyle,escapeinside={(*@}{@*)}]
(*@\mutantid{M0}@*) assign right = ~(scancode == 16'he074);

(*@\mutantid{M2}@*) assign up    = ~(scancode == 16'he075);

(*@\mutantid{M4}@*) assign down  = ~(scancode == 16'he072);

(*@\mutantid{M6}@*) assign left  = ~(scancode == 16'he06b);
\end{lstlisting}

\vspace{-6pt}


\mutheading{Stuck-at-high output faults}

\begin{lstlisting}[style=verilogstyle,escapeinside={(*@}{@*)}]
(*@\mutantid{M1}@*) assign left  = 1'h1;

(*@\mutantid{M5}@*) assign up    = 1'h1;

(*@\mutantid{M8}@*) assign right = 1'h1;
\end{lstlisting}

\vspace{-6pt}


\mutheading{Stuck-at-low output faults}

\begin{tcolorbox}[
  enhanced,
  colback=covrgreen!10,
  colframe=covrgreen!35,
  boxrule=0.2pt,
  arc=0.5mm,
  left=1pt,
  right=1pt,
  top=1pt,
  bottom=1pt,
  boxsep=0pt
]

\begin{lstlisting}[style=verilogstyle,escapeinside={(*@}{@*)}]
(*@\mutantid{M3}@*) assign down  = 1'h0;

(*@\mutantid{M7}@*) assign left  = 1'h0;

(*@\mutantid{M9}@*) assign right = 1'h0;
\end{lstlisting}

\end{tcolorbox}

\vspace{1pt}

{\tiny\color{black!65}
Green-highlighted mutants are killed by CovR but missed by Pro-V.
}

\end{tcolorbox}
\end{minipage}

\vspace{-4pt}

\begin{tcolorbox}[title={(d) Pro-V Stimulus}]
\centering
\scriptsize
\textbf{Mutation Score:} 70.0\% \qquad
\textbf{Killed:} 7/10

\vspace{2pt}

{\fontsize{5.8}{6.3}\selectfont
\setlength{\tabcolsep}{7pt}
\renewcommand{\arraystretch}{0.96}
\begin{tabular}{c|ccccc}
\textbf{Test} & 0 & 1 & 2 & 3 & 4 \\
\hline
\texttt{scancode} & 1f7c & f072 & 29f1 & 30e1 & fb17 \\
\texttt{left}     & 0 & 0 & 0 & 0 & 0 \\
\texttt{down}     & 0 & 0 & 0 & 0 & 0 \\
\texttt{right}    & 0 & 0 & 0 & 0 & 0 \\
\texttt{up}       & 0 & 0 & 0 & 0 & 0 \\
\end{tabular}
}

\vspace{3pt}

\raggedright
\scriptsize
\textbf{Killed Mutants:}
\killchip{M0} \killchip{M1} \killchip{M2} \killchip{M4}
\killchip{M5} \killchip{M6} \killchip{M8}

\textbf{Survived Mutants:}
\survivechip{M3} \survivechip{M7} \survivechip{M9}

\vspace{2pt}
\footnotesize
Most sampled scancodes are invalid, so the expected output remains all-zero.
This exposes faults that perturb invalid-code behavior, but misses
valid-code stuck-at-zero decode faults.
\end{tcolorbox}

\vspace{-3pt}

\begin{tcolorbox}[title={(e) CovR Stimulus}]
\centering
\scriptsize
\textbf{Mutation Score:} 100.0\% \qquad
\textbf{Killed:} 10/10

\vspace{2pt}

{\fontsize{6.8}{7.3}\selectfont
\setlength{\tabcolsep}{9pt}
\renewcommand{\arraystretch}{0.96}
\begin{tabular}{c|cccc}
\textbf{Test} & 0 & 1 & 2 & 3 \\
\hline
\texttt{scancode} & e06b & e072 & e074 & e075 \\
\texttt{left}
& \textcolor{covrgreen}{\textbf{1}} & 0 & 0 & 0 \\
\texttt{down}
& 0 & \textcolor{covrgreen}{\textbf{1}} & 0 & 0 \\
\texttt{right}
& 0 & 0 & \textcolor{covrgreen}{\textbf{1}} & 0 \\
\texttt{up}
& 0 & 0 & 0 & \textcolor{covrgreen}{\textbf{1}} \\
\end{tabular}
}

\vspace{3pt}

\raggedright
\scriptsize
\textbf{Killed Mutants:}
\killchip{M0} \killchip{M1} \killchip{M2} \killchip{M3}
\killchip{M4} \killchip{M5} \killchip{M6} \killchip{M7}
\killchip{M8} \killchip{M9}

\textbf{Survived Mutants:}
\survivechip{None}

\vspace{2pt}
\footnotesize
CovR exercises each semantic scancode, activating every decode path and
exposing both inverted and stuck-at-zero faults that survive under
invalid-only stimulus.
\end{tcolorbox}

\caption{
CovR improves mutation detection on
\texttt{Prob106\_always\_nolatches}.
}
\label{fig:covr-mutant-killing-scancode}
\end{figure}

\begin{figure}[H]
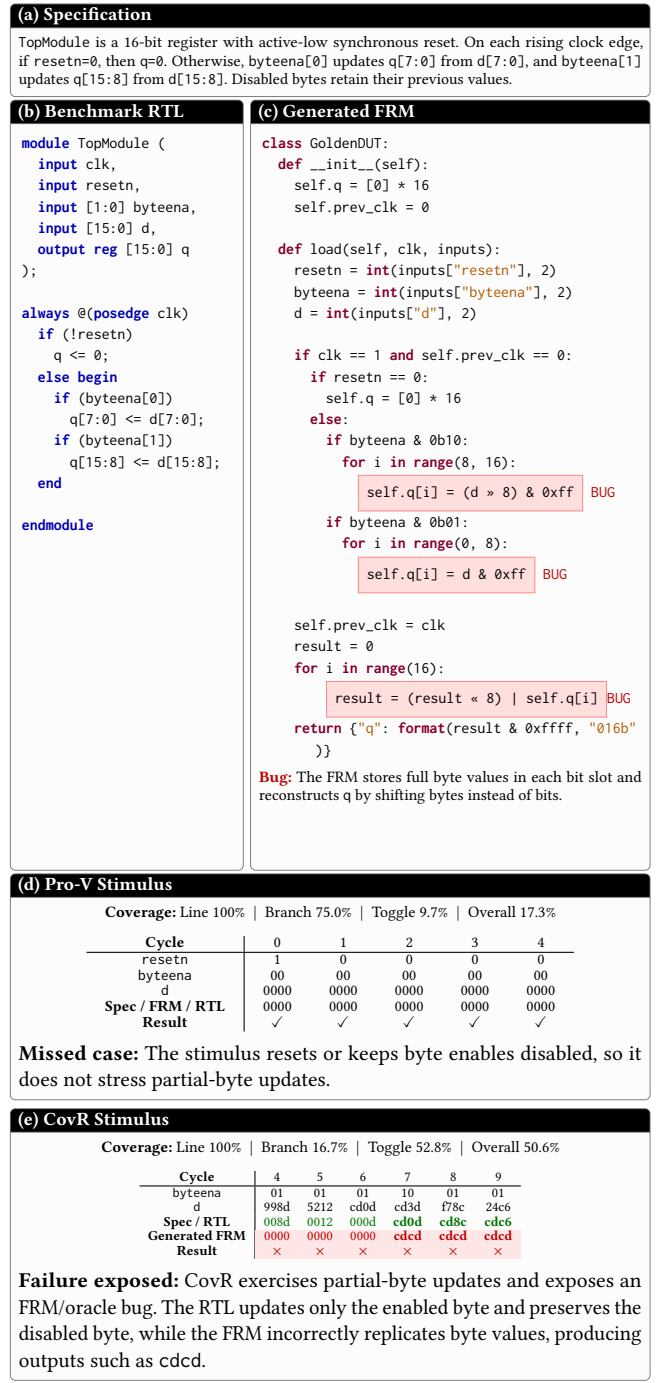

\centering
\scriptsize


\lstdefinestyle{verilogstyle}{
  language=Verilog,
  basicstyle=\ttfamily\scriptsize,
  keywordstyle=\color{blue!70!black}\bfseries,
  commentstyle=\color{green!40!black},
  stringstyle=\color{orange!70!black},
  stepnumber=1,
  numbersep=1pt,
  breaklines=true,
  columns=fullflexible,
  keepspaces=true,
  showstringspaces=false,
  frame=none,
  xleftmargin=0.2em,
  framexleftmargin=0.1em,
  aboveskip=1pt,
  belowskip=1pt
}

\lstdefinestyle{pythonstyle}{
  language=Python,
  basicstyle=\ttfamily\scriptsize,
  keywordstyle=\color{purple!70!black}\bfseries,
  commentstyle=\color{green!40!black},
  stringstyle=\color{orange!70!black},
  stepnumber=1,
  numbersep=1pt,
  breaklines=true,
  columns=fullflexible,
  keepspaces=true,
  showstringspaces=false,
  frame=none,
  xleftmargin=0.2em,
  framexleftmargin=0.1em,
  aboveskip=1pt,
  belowskip=1pt,
  escapeinside={(*@}{@*)}
}

\tcbset{
  enhanced,
  boxsep=1pt,
  left=2pt,
  right=2pt,
  top=2pt,
  bottom=2pt,
  arc=0.6mm,
  boxrule=0.35pt,
  colback=black!1,
  colframe=black!55,
  coltitle=white,
  colbacktitle=black,
  fonttitle=\bfseries\footnotesize,
  before skip=2pt,
  after skip=2pt
}

\begin{tcolorbox}[title={(a) Specification}]
\texttt{TopModule} is a 16-bit register with active-low synchronous reset.
On each rising clock edge, if \texttt{resetn=0}, then \texttt{q=0}.
Otherwise, \texttt{byteena[0]} updates \texttt{q[7:0]} from
\texttt{d[7:0]}, and \texttt{byteena[1]} updates \texttt{q[15:8]} from
\texttt{d[15:8]}. Disabled bytes retain their previous values.
\end{tcolorbox}

\vspace{-1pt}

\begin{minipage}[t]{0.365\columnwidth}
\begin{tcolorbox}[
  title={(b) Benchmark RTL},
  height=10.2cm,
  valign=top
]
\begin{lstlisting}[style=verilogstyle]
module TopModule (
  input clk,
  input resetn,
  input [1:0] byteena,
  input [15:0] d,
  output reg [15:0] q
);

always @(posedge clk)
  if (!resetn)
    q <= 0;
  else begin
    if (byteena[0])
      q[7:0] <= d[7:0];
    if (byteena[1])
      q[15:8] <= d[15:8];
  end

endmodule
\end{lstlisting}
\end{tcolorbox}
\end{minipage}
\hfill
\begin{minipage}[t]{0.625\columnwidth}
\begin{tcolorbox}[
  title={(c) Generated FRM},
  height=10.2cm,
  valign=top
]
\begin{lstlisting}[style=pythonstyle]
class GoldenDUT:
  def __init__(self):
    self.q = [0] * 16
    self.prev_clk = 0

  def load(self, clk, inputs):
    resetn = int(inputs["resetn"], 2)
    byteena = int(inputs["byteena"], 2)
    d = int(inputs["d"], 2)

    if clk == 1 and self.prev_clk == 0:
      if resetn == 0:
        self.q = [0] * 16
      else:
        if byteena & 0b10:
          for i in range(8, 16):
            (*@\fcolorbox{red!45}{red!13}{\strut\texttt{self.q[i] = (d >> 8) \& 0xff}}@*) (*@\textcolor{red!75!black}{\scriptsize BUG}@*)
        if byteena & 0b01:
          for i in range(0, 8):
            (*@\fcolorbox{red!45}{red!13}{\strut\texttt{self.q[i] = d \& 0xff}}@*) (*@\textcolor{red!75!black}{\scriptsize BUG}@*)

    self.prev_clk = clk
    result = 0
    for i in range(16):
        (*@\fcolorbox{red!45}{red!13}{\strut\texttt{result = (result << 8) | self.q[i]}}@*)(*@\textcolor{red!75!black}{\scriptsize BUG}@*)
    return {"q": format(result & 0xffff, "016b")}
\end{lstlisting}

\vspace{2pt}

{\scriptsize
\textcolor{red!75!black}{\textbf{Bug:}}
The FRM stores full byte values in each bit slot and reconstructs
\texttt{q} by shifting bytes instead of bits.}

\end{tcolorbox}
\end{minipage}

\vspace{-1pt}

\begin{tcolorbox}[
  title={(d) Pro-V Stimulus}
]

\begin{center}
\scriptsize
\textbf{Coverage:}
Line 100\% \;|\;
Branch 75.0\% \;|\;
Toggle 9.7\% \;|\;
Overall 17.3\%
\end{center}

\vspace{1pt}

\begin{center}
{\fontsize{5.8}{6.3}\selectfont
\setlength{\tabcolsep}{7pt}
\renewcommand{\arraystretch}{0.94}

\begin{tabular}{c|ccccc}
\textbf{Cycle} & 0 & 1 & 2 & 3 & 4 \\
\hline
\texttt{resetn}  & 1 & 0 & 0 & 0 & 0 \\
\texttt{byteena} & 00 & 00 & 00 & 00 & 00 \\
\texttt{d}       & 0000 & 0000 & 0000 & 0000 & 0000 \\
\textbf{Spec / FRM / RTL}
& 0000 & 0000 & 0000 & 0000 & 0000 \\
\textbf{Result}
& $\checkmark$ & $\checkmark$ & $\checkmark$ &
$\checkmark$ & $\checkmark$
\end{tabular}
}
\end{center}

\vspace{2pt}

\small
\textbf{Missed case:} The stimulus resets or keeps byte enables disabled, so
it does not stress partial-byte updates.

\end{tcolorbox}

\vspace{-1pt}

\begin{tcolorbox}[
  title={(e) CovR Stimulus}
]

\begin{center}
\scriptsize
\textbf{Coverage:}
Line 100\% \;|\;
Branch 16.7\% \;|\;
Toggle 52.8\% \;|\;
Overall 50.6\%
\end{center}

\vspace{1pt}

\begin{center}
{\fontsize{5.2}{5.8}\selectfont
\setlength{\tabcolsep}{3.2pt}
\renewcommand{\arraystretch}{0.94}

\begin{tabular}{c|cccccc}
\textbf{Cycle}
& 4 & 5 & 6 & 7 & 8 & 9 \\
\hline

\texttt{byteena}
& 01 & 01 & 01 & 10 & 01 & 01 \\

\texttt{d}
& 998d & 5212 & cd0d & cd3d & f78c & 24c6 \\

\textbf{Spec / RTL}
& \textcolor{green!45!black}{008d}
& \textcolor{green!45!black}{0012}
& \textcolor{green!45!black}{000d}
& \textcolor{green!45!black}{\textbf{cd0d}}
& \textcolor{green!45!black}{\textbf{cd8c}}
& \textcolor{green!45!black}{\textbf{cdc6}} \\

\textbf{Generated FRM}
& \cellcolor{red!10}\textcolor{red!75!black}{0000}
& \cellcolor{red!10}\textcolor{red!75!black}{0000}
& \cellcolor{red!10}\textcolor{red!75!black}{0000}
& \cellcolor{red!10}\textcolor{red!75!black}{\textbf{cdcd}}
& \cellcolor{red!10}\textcolor{red!75!black}{\textbf{cdcd}}
& \cellcolor{red!10}\textcolor{red!75!black}{\textbf{cdcd}} \\

\textbf{Result}
& \cellcolor{red!10}\textcolor{red!75!black}{$\times$}
& \cellcolor{red!10}\textcolor{red!75!black}{$\times$}
& \cellcolor{red!10}\textcolor{red!75!black}{$\times$}
& \cellcolor{red!10}\textcolor{red!75!black}{$\times$}
& \cellcolor{red!10}\textcolor{red!75!black}{$\times$}
& \cellcolor{red!10}\textcolor{red!75!black}{$\times$}

\end{tabular}
}
\end{center}

\vspace{2pt}

\small
\textbf{Failure exposed:} CovR exercises partial-byte updates and exposes an
FRM/oracle bug. The RTL updates only the enabled byte and preserves the
disabled byte, while the FRM incorrectly replicates byte values, producing
outputs such as \texttt{cdcd}.

\end{tcolorbox}

\vspace{-3pt}

\caption{
CovR reveals an oracle bug in the generated FRM for a byte-enable register.
The RTL correctly updates only enabled bytes, while the FRM incorrectly
replicates byte values across its internal state.
}
\label{fig:covr-reveals-byte-enable-frm-bug}

\end{figure}

Figure~\ref{fig:covr-reveals-byte-enable-frm-bug} shows how CovR’s broader stimulus exploration uncovers latent failures in the FRM generated by PRO-V-R1~\cite{zhao2025pro}.

\begin{figure}[H]
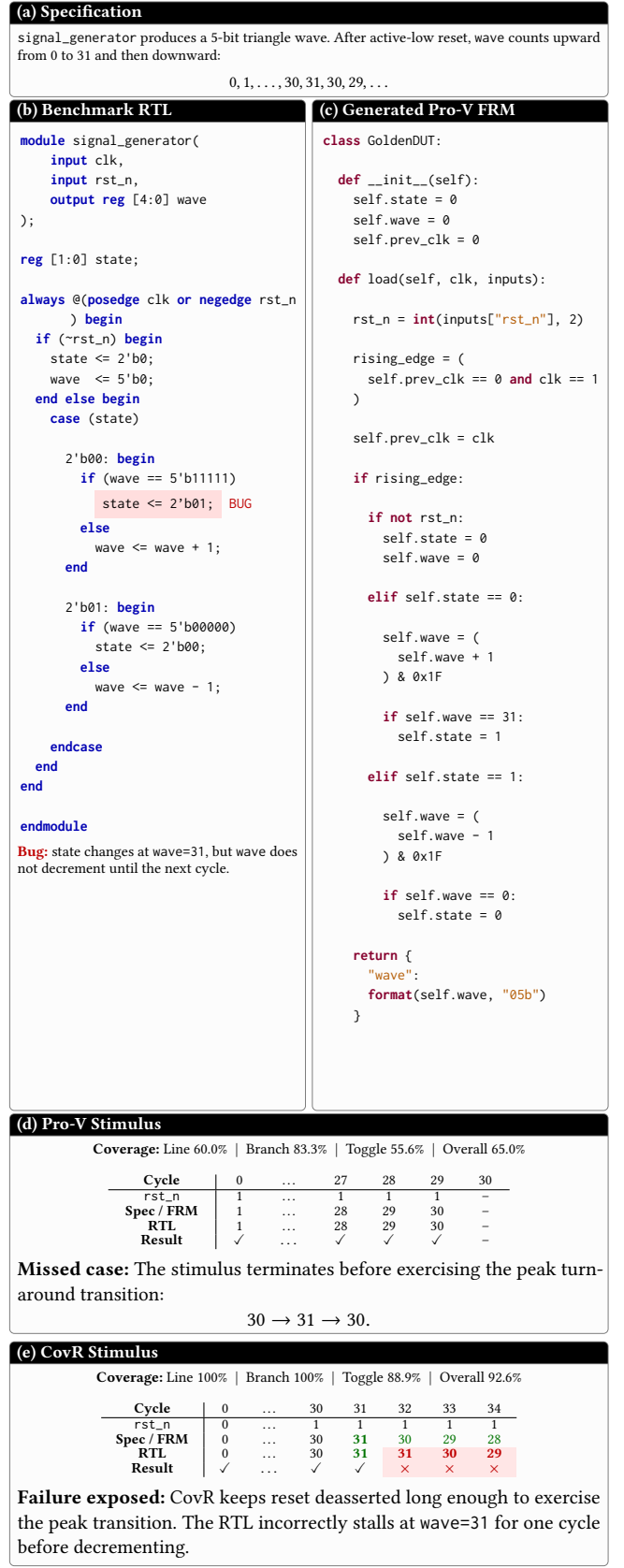

\centering
\scriptsize


\lstdefinestyle{verilogstyle}{
  language=Verilog,
  basicstyle=\ttfamily\scriptsize,
  keywordstyle=\color{blue!70!black}\bfseries,
  commentstyle=\color{green!40!black},
  stringstyle=\color{orange!70!black},
  stepnumber=1,
  numbersep=1pt,
  breaklines=true,
  columns=fullflexible,
  keepspaces=true,
  showstringspaces=false,
  frame=none,
  xleftmargin=0.2em,
  framexleftmargin=0.1em,
  aboveskip=1pt,
  belowskip=1pt,
  escapeinside={(*@}{@*)}
}

\lstdefinestyle{pythonstyle}{
  language=Python,
  basicstyle=\ttfamily\scriptsize,
  keywordstyle=\color{purple!70!black}\bfseries,
  commentstyle=\color{green!40!black},
  stringstyle=\color{orange!70!black},
  stepnumber=1,
  numbersep=1pt,
  breaklines=true,
  columns=fullflexible,
  keepspaces=true,
  showstringspaces=false,
  frame=none,
  xleftmargin=0.2em,
  framexleftmargin=0.1em,
  aboveskip=1pt,
  belowskip=1pt
}

\tcbset{
  enhanced,
  boxsep=1pt,
  left=2pt,
  right=2pt,
  top=2pt,
  bottom=2pt,
  arc=0.6mm,
  boxrule=0.35pt,
  colback=black!1,
  colframe=black!55,
  coltitle=white,
  colbacktitle=black,
  fonttitle=\bfseries\footnotesize,
  before skip=2pt,
  after skip=2pt
}

\begin{tcolorbox}[title={(a) Specification}]
\texttt{signal\_generator} produces a 5-bit triangle wave. After active-low
reset, \texttt{wave} counts upward from 0 to 31 and then downward:
\[
0,1,\ldots,30,31,30,29,\ldots
\]
\end{tcolorbox}

\vspace{-1pt}

\begin{minipage}[t]{0.495\columnwidth}
\begin{tcolorbox}[
  title={(b) Benchmark RTL},
  height=14.3cm,
  valign=top
]

\begin{lstlisting}[style=verilogstyle]
module signal_generator(
    input clk,
    input rst_n,
    output reg [4:0] wave
);

reg [1:0] state;

always @(posedge clk or negedge rst_n) begin
  if (~rst_n) begin
    state <= 2'b0;
    wave  <= 5'b0;
  end else begin
    case (state)

      2'b00: begin
        if (wave == 5'b11111)
          (*@\colorbox{red!13}{state <= 2'b01;}@*) (*@\textcolor{red!75!black}{\scriptsize BUG}@*)
        else
          wave <= wave + 1;
      end

      2'b01: begin
        if (wave == 5'b00000)
          state <= 2'b00;
        else
          wave <= wave - 1;
      end

    endcase
  end
end

endmodule
\end{lstlisting}

\vspace{2pt}
{\scriptsize\textcolor{red!75!black}{\textbf{Bug:}} state changes at
\texttt{wave=31}, but \texttt{wave} does not decrement until the next cycle.}

\end{tcolorbox}
\end{minipage}
\hfill
\begin{minipage}[t]{0.495\columnwidth}
\begin{tcolorbox}[
  title={(c) Generated Pro-V FRM},
  height=14.3cm,
  valign=top
]

\begin{lstlisting}[style=pythonstyle]
class GoldenDUT:

  def __init__(self):
    self.state = 0
    self.wave = 0
    self.prev_clk = 0

  def load(self, clk, inputs):

    rst_n = int(inputs["rst_n"], 2)

    rising_edge = (
      self.prev_clk == 0 and clk == 1
    )

    self.prev_clk = clk

    if rising_edge:

      if not rst_n:
        self.state = 0
        self.wave = 0

      elif self.state == 0:

        self.wave = (
          self.wave + 1
        ) & 0x1F

        if self.wave == 31:
          self.state = 1

      elif self.state == 1:

        self.wave = (
          self.wave - 1
        ) & 0x1F

        if self.wave == 0:
          self.state = 0

    return {
      "wave":
      format(self.wave, "05b")
    }
\end{lstlisting}

\end{tcolorbox}
\end{minipage}

\vspace{-1pt}

\begin{tcolorbox}[title={(d) Pro-V Stimulus}]
\begin{center}
\scriptsize
\textbf{Coverage:}
Line 60.0\% \;|\;
Branch 83.3\% \;|\;
Toggle 55.6\% \;|\;
Overall 65.0\%
\end{center}

\vspace{2pt}

\begin{center}
{\fontsize{5.8}{6.3}\selectfont
\setlength{\tabcolsep}{7pt}
\renewcommand{\arraystretch}{0.94}
\begin{tabular}{c|cccccc}
\textbf{Cycle}
& 0 & \dots & 27 & 28 & 29 & 30 \\
\hline
\texttt{rst\_n}
& 1 & \dots & 1 & 1 & 1 & -- \\
\textbf{Spec / FRM}
& 1 & \dots & 28 & 29 & 30 & -- \\
\textbf{RTL}
& 1 & \dots & 28 & 29 & 30 & -- \\
\textbf{Result}
& $\checkmark$ & $\dots$ & $\checkmark$ & $\checkmark$ & $\checkmark$ & -- \\
\end{tabular}
}
\end{center}

\vspace{2pt}

\small
\textbf{Missed case:} The stimulus terminates before exercising the peak
turn-around transition:
\[
30 \rightarrow 31 \rightarrow 30 .
\]
\end{tcolorbox}

\vspace{-1pt}

\begin{tcolorbox}[title={(e) CovR Stimulus}]
\begin{center}
\scriptsize
\textbf{Coverage:}
Line 100\% \;|\;
Branch 100\% \;|\;
Toggle 88.9\% \;|\;
Overall 92.6\%
\end{center}

\vspace{2pt}

\begin{center}
{\fontsize{5.8}{6.3}\selectfont
\setlength{\tabcolsep}{6pt}
\renewcommand{\arraystretch}{0.94}
\begin{tabular}{c|ccccccc}
\textbf{Cycle}
& 0 & \dots & 30 & 31 & 32 & 33 & 34 \\
\hline
\texttt{rst\_n}
& 0 & \dots & 1 & 1 & 1 & 1 & 1 \\
\textbf{Spec / FRM}
& 0 & \dots & 30
& \textcolor{green!45!black}{\textbf{31}}
& \textcolor{green!45!black}{30}
& \textcolor{green!45!black}{29}
& \textcolor{green!45!black}{28} \\
\textbf{RTL}
& 0 & \dots & 30
& \textcolor{green!45!black}{\textbf{31}}
& \cellcolor{red!10}\textcolor{red!75!black}{\textbf{31}}
& \cellcolor{red!10}\textcolor{red!75!black}{\textbf{30}}
& \cellcolor{red!10}\textcolor{red!75!black}{\textbf{29}} \\
\textbf{Result}
& $\checkmark$
& $\dots$
& $\checkmark$
& $\checkmark$
& \cellcolor{red!10}\textcolor{red!75!black}{$\times$}
& \cellcolor{red!10}\textcolor{red!75!black}{$\times$}
& \cellcolor{red!10}\textcolor{red!75!black}{$\times$}
\end{tabular}
}
\end{center}

\vspace{2pt}

\small
\textbf{Failure exposed:} CovR keeps reset deasserted long enough to exercise
the peak transition. The RTL incorrectly stalls at \texttt{wave=31}
for one cycle before decrementing.
\end{tcolorbox}

\vspace{-3pt}

\caption{
CovR generates longer temporal stimuli that reach the peak turn-around case
and expose an RTL bug missed by shorter static stimulus sequences.
}

\label{fig:covr-reveals-signal-generator-bug}
\end{figure}

Figure~\ref{fig:covr-reveals-signal-generator-bug} shows how CovR reveals a DUT RTL bug by exercising an uncovered edge case. 

\begin{figure}[H]
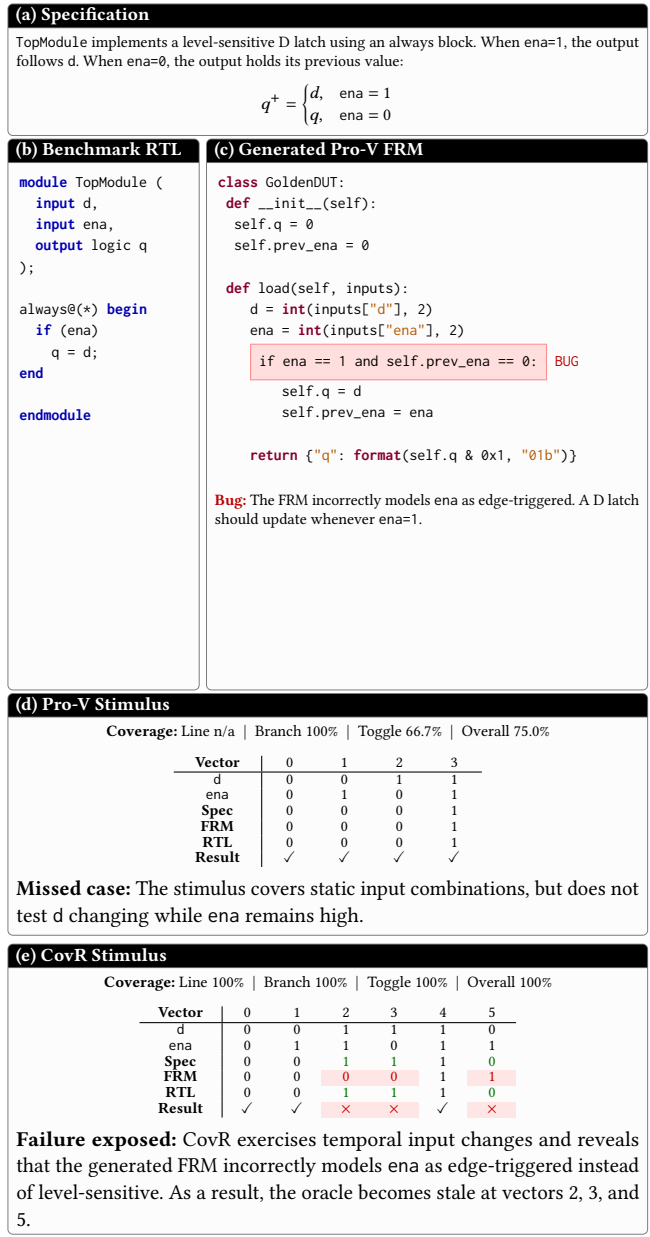

\centering
\scriptsize


\lstdefinestyle{verilogstyle}{
  language=Verilog,
  basicstyle=\ttfamily\scriptsize,
  keywordstyle=\color{blue!70!black}\bfseries,
  commentstyle=\color{green!40!black},
  stringstyle=\color{orange!70!black},
  stepnumber=1,
  numbersep=1pt,
  breaklines=true,
  columns=fullflexible,
  keepspaces=true,
  showstringspaces=false,
  frame=none,
  xleftmargin=0.2em,
  framexleftmargin=0.1em,
  aboveskip=1pt,
  belowskip=1pt
}

\lstdefinestyle{pythonstyle}{
  language=Python,
  basicstyle=\ttfamily\scriptsize,
  keywordstyle=\color{purple!70!black}\bfseries,
  commentstyle=\color{green!40!black},
  stringstyle=\color{orange!70!black},
  stepnumber=1,
  numbersep=1pt,
  breaklines=true,
  columns=fullflexible,
  keepspaces=true,
  showstringspaces=false,
  frame=none,
  xleftmargin=0.2em,
  framexleftmargin=0.1em,
  aboveskip=1pt,
  belowskip=1pt,
escapeinside={(*@}{@*)}
}

\tcbset{
  enhanced,
  boxsep=1pt,
  left=2pt,
  right=2pt,
  top=2pt,
  bottom=2pt,
  arc=0.6mm,
  boxrule=0.35pt,
  colback=black!1,
  colframe=black!55,
  coltitle=white,
  colbacktitle=black,
  fonttitle=\bfseries\footnotesize,
  before skip=2pt,
  after skip=2pt
}

\begin{tcolorbox}[title={(a) Specification}]
\texttt{TopModule} implements a level-sensitive D latch using an always block.
When \texttt{ena=1}, the output follows \texttt{d}. When \texttt{ena=0},
the output holds its previous value:
\[
q^{+} =
\begin{cases}
d, & \texttt{ena}=1 \\
q, & \texttt{ena}=0
\end{cases}
\]
\end{tcolorbox}

\vspace{-1pt}

\begin{minipage}[t]{0.3\columnwidth}
\begin{tcolorbox}[
  title={(b) Benchmark RTL},
  height=7.3cm,
  valign=top
]
\begin{lstlisting}[style=verilogstyle]
module TopModule (
  input d,
  input ena,
  output logic q
);

always@(*) begin
  if (ena)
    q = d;
end

endmodule
\end{lstlisting}
\end{tcolorbox}
\end{minipage}
\hfill
\begin{minipage}[t]{0.69\columnwidth}
\begin{tcolorbox}[
  title={(c) Generated Pro-V FRM},
  height=7.3cm,
  valign=top
]
\begin{lstlisting}[style=pythonstyle]
class GoldenDUT:
 def __init__(self):
  self.q = 0
  self.prev_ena = 0

 def load(self, inputs):
    d = int(inputs["d"], 2)
    ena = int(inputs["ena"], 2)
    (*@\fcolorbox{red!45}{red!13}{\strut\texttt{if ena == 1 and self.prev\_ena == 0:}}@*) (*@\textcolor{red!75!black}{\scriptsize BUG}@*)
        self.q = d
        self.prev_ena = ena
    
    return {"q": format(self.q & 0x1, "01b")}
\end{lstlisting}

\vspace{2pt}

{\scriptsize
\textcolor{red!75!black}{
\\
\textbf{Bug:}}
The FRM incorrectly models \texttt{ena} as edge-triggered.
A D latch should update whenever \texttt{ena=1}.}

\end{tcolorbox}
\end{minipage}

\vspace{-1pt}

\begin{tcolorbox}[
  title={(d) Pro-V Stimulus}
]

\begin{center}
\scriptsize
\textbf{Coverage:}
Line n/a \;|\;
Branch 100\% \;|\;
Toggle 66.7\% \;|\;
Overall 75.0\%
\end{center}

\vspace{1pt}

\begin{center}
{\fontsize{5.8}{6.3}\selectfont
\setlength{\tabcolsep}{8pt}
\renewcommand{\arraystretch}{0.94}

\begin{tabular}{c|cccc}
\textbf{Vector} & 0 & 1 & 2 & 3 \\
\hline
\texttt{d}      & 0 & 0 & 1 & 1 \\
\texttt{ena}    & 0 & 1 & 0 & 1 \\

\textbf{Spec}   &
0 & 0 & 0 & 1 \\

\textbf{FRM}    &
0 & 0 & 0 & 1 \\

\textbf{RTL}    &
0 & 0 & 0 & 1 \\

\textbf{Result} &
$\checkmark$ &
$\checkmark$ &
$\checkmark$ &
$\checkmark$
\end{tabular}
}
\end{center}

\vspace{2pt}

\small
\textbf{Missed case:} The stimulus covers static input combinations, but does
not test \texttt{d} changing while \texttt{ena} remains high.

\end{tcolorbox}

\vspace{-1pt}

\begin{tcolorbox}[
  title={(e) CovR Stimulus}
]

\begin{center}
\scriptsize
\textbf{Coverage:}
Line 100\% \;|\;
Branch 100\% \;|\;
Toggle 100\% \;|\;
Overall 100\%
\end{center}

\vspace{1pt}

\begin{center}
{\fontsize{5.8}{6.3}\selectfont
\setlength{\tabcolsep}{7pt}
\renewcommand{\arraystretch}{0.94}

\begin{tabular}{c|cccccc}
\textbf{Vector} & 0 & 1 & 2 & 3 & 4 & 5 \\
\hline

\texttt{d}      &
0 & 0 & 1 & 1 & 1 & 0 \\

\texttt{ena}    &
0 & 1 & 1 & 0 & 1 & 1 \\

\textbf{Spec} &
0 & 0 &
\textcolor{green!45!black}{1} &
\textcolor{green!45!black}{1} &
1 &
\textcolor{green!45!black}{0} \\

\textbf{FRM} &
0 & 0 &
\cellcolor{red!10}\textcolor{red!75!black}{0} &
\cellcolor{red!10}\textcolor{red!75!black}{0} &
1 &
\cellcolor{red!10}\textcolor{red!75!black}{1} \\

\textbf{RTL} &
0 & 0 &
\textcolor{green!45!black}{1} &
\textcolor{green!45!black}{1} &
1 &
\textcolor{green!45!black}{0} \\

\textbf{Result} &
$\checkmark$ &
$\checkmark$ &
\cellcolor{red!10}\textcolor{red!75!black}{$\times$} &
\cellcolor{red!10}\textcolor{red!75!black}{$\times$} &
$\checkmark$ &
\cellcolor{red!10}\textcolor{red!75!black}{$\times$}
\end{tabular}
}
\end{center}

\vspace{2pt}

\small
\textbf{Failure exposed:} CovR exercises temporal input changes and reveals
that the generated FRM incorrectly models \texttt{ena} as edge-triggered
instead of level-sensitive. As a result, the oracle becomes stale at
vectors 2, 3, and 5.

\end{tcolorbox}

\vspace{-3pt}

\caption{
CovR generates temporally diverse stimuli that expose an oracle bug in the
generated functional reference model, while static input-combination stimuli
fail to reveal the mismatch.
}

\label{fig:covr-reveals-dlatch-frm-bug}

\end{figure}

Figure~\ref{fig:covr-reveals-dlatch-frm-bug} illustrates an oracle bug in the generated FRM that remains hidden under PRO-V stimuli due to insufficient temporal input variation.


\begin{figure}[H]
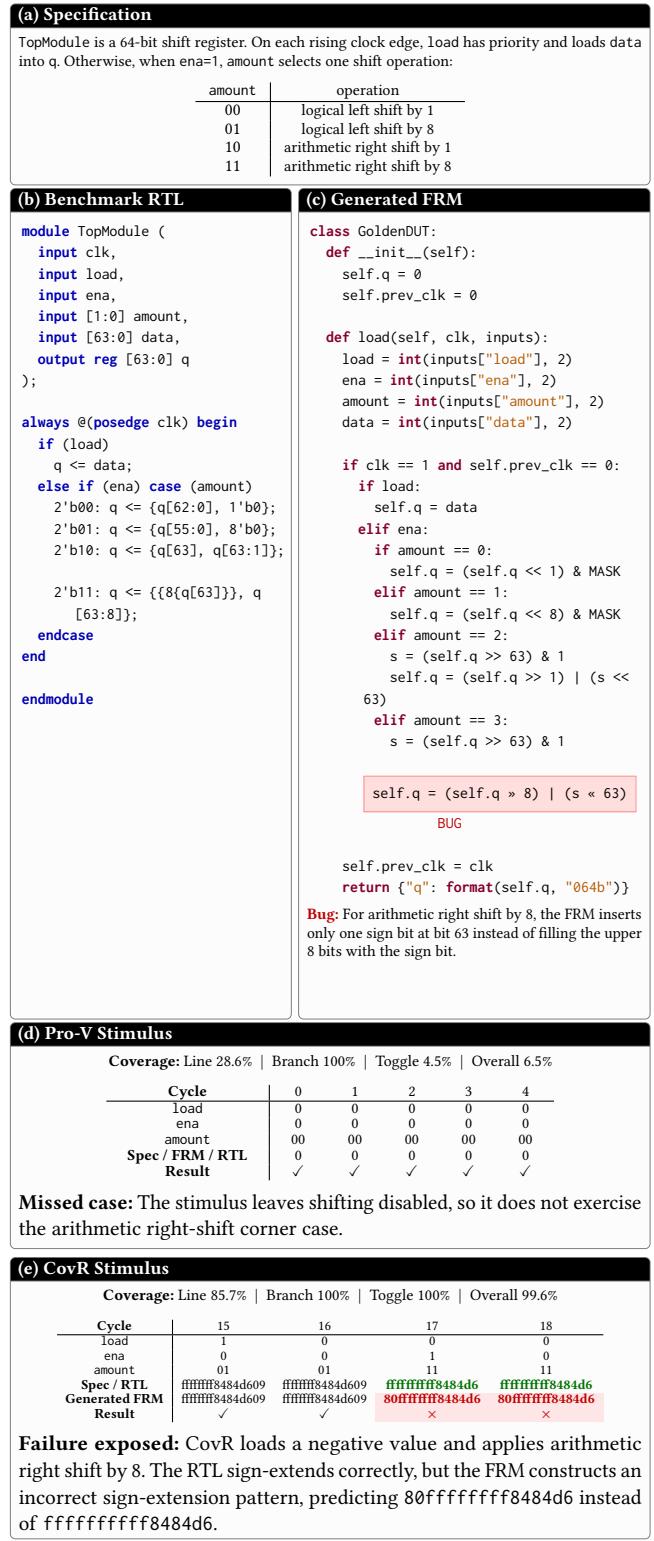

\centering
\scriptsize


\lstdefinestyle{verilogstyle}{
  language=Verilog,
  basicstyle=\ttfamily\scriptsize,
  keywordstyle=\color{blue!70!black}\bfseries,
  commentstyle=\color{green!40!black},
  stringstyle=\color{orange!70!black},
  stepnumber=1,
  numbersep=1pt,
  breaklines=true,
  columns=fullflexible,
  keepspaces=true,
  showstringspaces=false,
  frame=none,
  xleftmargin=0.2em,
  framexleftmargin=0.1em,
  aboveskip=1pt,
  belowskip=1pt
}

\lstdefinestyle{pythonstyle}{
  language=Python,
  basicstyle=\ttfamily\scriptsize,
  keywordstyle=\color{purple!70!black}\bfseries,
  commentstyle=\color{green!40!black},
  stringstyle=\color{orange!70!black},
  stepnumber=1,
  numbersep=1pt,
  breaklines=true,
  columns=fullflexible,
  keepspaces=true,
  showstringspaces=false,
  frame=none,
  xleftmargin=0.2em,
  framexleftmargin=0.1em,
  aboveskip=1pt,
  belowskip=1pt,
  escapeinside={(*@}{@*)}
}

\tcbset{
  enhanced,
  boxsep=1pt,
  left=2pt,
  right=2pt,
  top=2pt,
  bottom=2pt,
  arc=0.6mm,
  boxrule=0.35pt,
  colback=black!1,
  colframe=black!55,
  coltitle=white,
  colbacktitle=black,
  fonttitle=\bfseries\footnotesize,
  before skip=2pt,
  after skip=2pt
}

\begin{tcolorbox}[title={(a) Specification}]
\texttt{TopModule} is a 64-bit shift register. On each rising clock edge,
\texttt{load} has priority and loads \texttt{data} into \texttt{q}. Otherwise,
when \texttt{ena=1}, \texttt{amount} selects one shift operation:
\[
\begin{array}{c|c}
\texttt{amount} & \text{operation} \\
\hline
00 & \text{logical left shift by 1} \\
01 & \text{logical left shift by 8} \\
10 & \text{arithmetic right shift by 1} \\
11 & \text{arithmetic right shift by 8}
\end{array}
\]
\end{tcolorbox}

\vspace{-1pt}

\begin{minipage}[t]{0.44\columnwidth}
\begin{tcolorbox}[
  title={(b) Benchmark RTL},
  height=11.0cm,
  valign=top
]
\begin{lstlisting}[style=verilogstyle]
module TopModule (
  input clk,
  input load,
  input ena,
  input [1:0] amount,
  input [63:0] data,
  output reg [63:0] q
);

always @(posedge clk) begin
  if (load)
    q <= data;
  else if (ena) case (amount)
    2'b00: q <= {q[62:0], 1'b0};
    2'b01: q <= {q[55:0], 8'b0};
    2'b10: q <= {q[63], q[63:1]};
    2'b11: q <= {{8{q[63]}}, q[63:8]};
  endcase
end

endmodule
\end{lstlisting}
\end{tcolorbox}
\end{minipage}
\hfill
\begin{minipage}[t]{0.55\columnwidth}
\begin{tcolorbox}[
  title={(c) Generated FRM},
  height=11.0cm,
  valign=top
]
\begin{lstlisting}[style=pythonstyle]
class GoldenDUT:
  def __init__(self):
    self.q = 0
    self.prev_clk = 0

  def load(self, clk, inputs):
    load = int(inputs["load"], 2)
    ena = int(inputs["ena"], 2)
    amount = int(inputs["amount"], 2)
    data = int(inputs["data"], 2)

    if clk == 1 and self.prev_clk == 0:
      if load:
        self.q = data
      elif ena:
        if amount == 0:
          self.q = (self.q << 1) & MASK
        elif amount == 1:
          self.q = (self.q << 8) & MASK
        elif amount == 2:
          s = (self.q >> 63) & 1
          self.q = (self.q >> 1) | (s << 63)
        elif amount == 3:
          s = (self.q >> 63) & 1
                    (*@\fcolorbox{red!45}{red!13}{\strut\texttt{self.q = (self.q >> 8) | (s << 63)}}@*)
                (*@\textcolor{red!75!black}{\scriptsize BUG}@*)
            
    self.prev_clk = clk
    return {"q": format(self.q, "064b")}
\end{lstlisting}

\vspace{2pt}

{\scriptsize
\textcolor{red!75!black}{\textbf{Bug:}}
For arithmetic right shift by 8, the FRM inserts only one sign bit at bit 63
instead of filling the upper 8 bits with the sign bit.}

\end{tcolorbox}
\end{minipage}

\vspace{-1pt}

\begin{tcolorbox}[
  title={(d) Pro-V Stimulus}
]

\begin{center}
\scriptsize
\textbf{Coverage:}
Line 28.6\% \;|\;
Branch 100\% \;|\;
Toggle 4.5\% \;|\;
Overall 6.5\%
\end{center}

\vspace{1pt}

\begin{center}
{\fontsize{5.8}{6.3}\selectfont
\setlength{\tabcolsep}{8pt}
\renewcommand{\arraystretch}{0.94}

\begin{tabular}{c|ccccc}
\textbf{Cycle} & 0 & 1 & 2 & 3 & 4 \\
\hline
\texttt{load}   & 0 & 0 & 0 & 0 & 0 \\
\texttt{ena}    & 0 & 0 & 0 & 0 & 0 \\
\texttt{amount} & 00 & 00 & 00 & 00 & 00 \\
\textbf{Spec / FRM / RTL}
& 0 & 0 & 0 & 0 & 0 \\
\textbf{Result}
& $\checkmark$ & $\checkmark$ & $\checkmark$ &
  $\checkmark$ & $\checkmark$
\end{tabular}
}
\end{center}

\vspace{2pt}

\small
\textbf{Missed case:} The stimulus leaves shifting disabled, so it does not
exercise the arithmetic right-shift corner case.

\end{tcolorbox}

\vspace{-1pt}

\begin{tcolorbox}[
  title={(e) CovR Stimulus}
]

\begin{center}
\scriptsize
\textbf{Coverage:}
Line 85.7\% \;|\;
Branch 100\% \;|\;
Toggle 100\% \;|\;
Overall 99.6\%
\end{center}

\vspace{1pt}

\begin{center}
{\fontsize{5.2}{5.8}\selectfont
\setlength{\tabcolsep}{3.2pt}
\renewcommand{\arraystretch}{0.94}

\begin{tabular}{c|cccc}
\textbf{Cycle} & 15 & 16 & 17 & 18 \\
\hline
\texttt{load}   & 1 & 0 & 0 & 0 \\
\texttt{ena}    & 0 & 0 & 1 & 0 \\
\texttt{amount} & 01 & 01 & 11 & 11 \\
\textbf{Spec / RTL}
& ffffffff8484d609
& ffffffff8484d609
& \textcolor{green!45!black}{\textbf{ffffffffff8484d6}}
& \textcolor{green!45!black}{\textbf{ffffffffff8484d6}} \\
\textbf{Generated FRM}
& ffffffff8484d609
& ffffffff8484d609
& \cellcolor{red!10}\textcolor{red!75!black}{\textbf{80ffffffff8484d6}}
& \cellcolor{red!10}\textcolor{red!75!black}{\textbf{80ffffffff8484d6}} \\
\textbf{Result}
& $\checkmark$
& $\checkmark$
& \cellcolor{red!10}\textcolor{red!75!black}{$\times$}
& \cellcolor{red!10}\textcolor{red!75!black}{$\times$}
\end{tabular}
}
\end{center}

\vspace{2pt}

\small
\textbf{Failure exposed:} CovR loads a negative value and applies arithmetic
right shift by 8. The RTL sign-extends correctly, but the FRM constructs an
incorrect sign-extension pattern, predicting \texttt{80ffffffff8484d6}
instead of \texttt{ffffffffff8484d6}.

\end{tcolorbox}

\vspace{-3pt}

\caption{
CovR reveals an oracle bug in the generated FRM for
\texttt{Prob115\_shift18}. The RTL correctly sign-extends during arithmetic
right shift by 8, while the FRM constructs an incorrect sign-extension pattern.
}
\label{fig:covr-reveals-shift18-frm-bug}

\end{figure}

\section{CovR Stimuli Improves Mutation Detection Score}

In this section, we present two case studies on how CovR's generated stimulus improves mutation detection score. 


\begin{figure}[H]
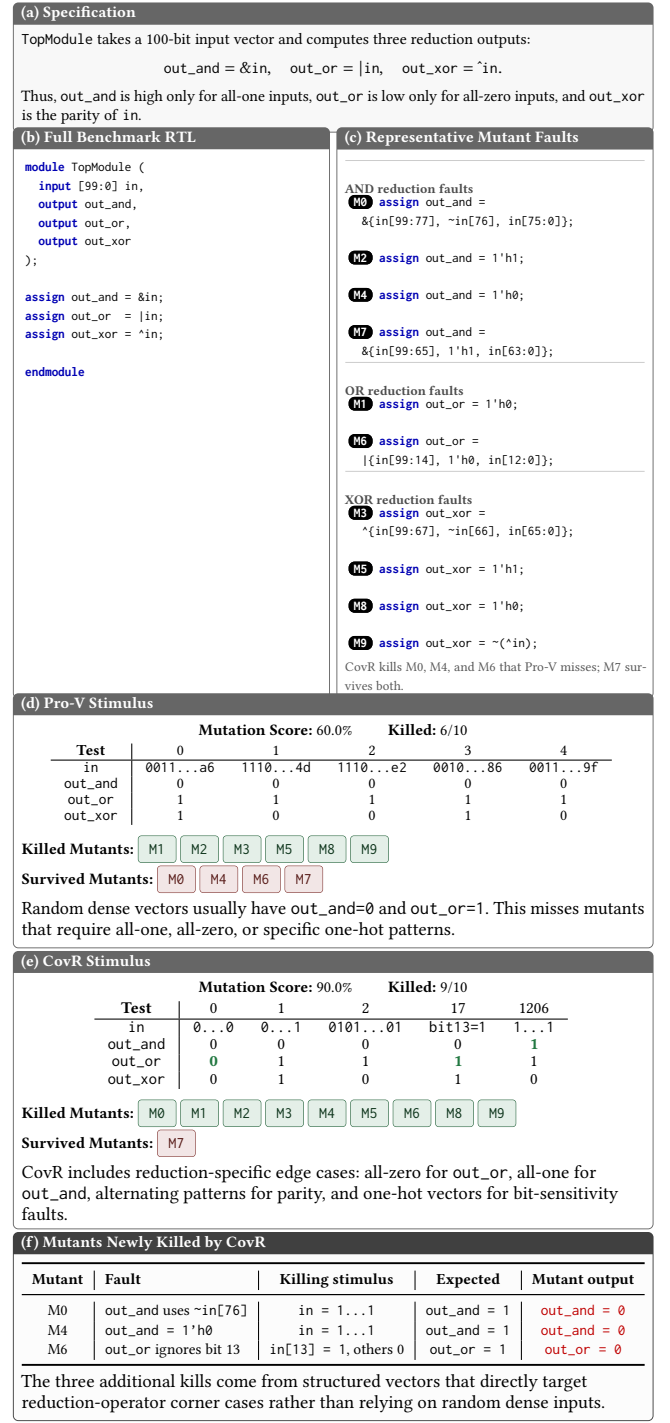

\centering
\scriptsize

\definecolor{covrgreen}{RGB}{34,120,65}
\definecolor{provred}{RGB}{150,50,45}
\definecolor{panelgray}{RGB}{248,248,248}
\definecolor{misshighlight}{RGB}{238,238,238}

\newtcbox{\mutantid}{
  on line,
  colback=black,
  colframe=black,
  coltext=white,
  boxrule=0pt,
  arc=0.8mm,
  left=2pt,
  right=2pt,
  top=1pt,
  bottom=1pt,
  boxsep=0pt,
  fontupper=\bfseries\tiny\ttfamily
}

\newcommand{\killchip}[1]{%
  \tcbox[
    on line,
    colback=covrgreen!12,
    colframe=covrgreen!75!black,
    coltext=covrgreen!45!black,
    boxrule=0.2pt,
    arc=0.4mm,
    left=1.2pt,right=1.2pt,top=0.2pt,bottom=0.2pt
  ]{\scriptsize\ttfamily #1}%
}

\newcommand{\survivechip}[1]{%
  \tcbox[
    on line,
    colback=provred!12,
    colframe=provred!80!black,
    coltext=provred!60!black,
    boxrule=0.2pt,
    arc=0.4mm,
    left=1.2pt,right=1.2pt,top=0.2pt,bottom=0.2pt
  ]{\scriptsize\ttfamily #1}%
}

\newcommand{\mutheading}[1]{%
\vspace{1pt}
{\color{black!20}\rule{\linewidth}{0.25pt}}
\vspace{-2pt}

{\bfseries\tiny\color{black!70} #1}
\vspace{-2pt}
}

\lstdefinestyle{verilogstyle}{
  language=Verilog,
  basicstyle=\ttfamily\tiny,
  keywordstyle=\color{blue!70!black}\bfseries,
  commentstyle=\color{green!40!black},
  stringstyle=\color{orange!70!black},
  breaklines=true,
  columns=fullflexible,
  keepspaces=true,
  showstringspaces=false,
  frame=none,
  xleftmargin=0.25em,
  aboveskip=1pt,
  belowskip=1pt
}

\tcbset{
  enhanced,
  boxsep=1pt,
  left=2pt,
  right=2pt,
  top=2pt,
  bottom=2pt,
  arc=0.6mm,
  boxrule=0.35pt,
  colframe=black!60,
  colback=white,
  fonttitle=\bfseries\scriptsize,
  before skip=2pt,
  after skip=2pt
}

\begin{tcolorbox}[title={(a) Specification}, colback=panelgray]
\texttt{TopModule} takes a 100-bit input vector and computes three reduction
outputs:
\[
\texttt{out\_and}=\&\texttt{in},\quad
\texttt{out\_or}=|\texttt{in},\quad
\texttt{out\_xor}=\hat{\ }\texttt{in}.
\]
Thus, \texttt{out\_and} is high only for all-one inputs, \texttt{out\_or}
is low only for all-zero inputs, and \texttt{out\_xor} is the parity of
\texttt{in}.
\end{tcolorbox}

\vspace{-4pt}

\begin{minipage}[t]{0.495\columnwidth}
\begin{tcolorbox}[title={(b) Full Benchmark RTL}, equal height group=A]
\begin{lstlisting}[style=verilogstyle]
module TopModule (
  input [99:0] in,
  output out_and,
  output out_or,
  output out_xor
);

assign out_and = &in;
assign out_or  = |in;
assign out_xor = ^in;

endmodule
\end{lstlisting}
\end{tcolorbox}
\end{minipage}
\hfill
\begin{minipage}[t]{0.495\columnwidth}
\begin{tcolorbox}[
  title={(c) Representative Mutant Faults},
  equal height group=A
]

\mutheading{AND reduction faults}

\begin{lstlisting}[style=verilogstyle,escapeinside={(*@}{@*)}]
(*@\mutantid{M0}@*) assign out_and =
  &{in[99:77], ~in[76], in[75:0]};

(*@\mutantid{M2}@*) assign out_and = 1'h1;

(*@\mutantid{M4}@*) assign out_and = 1'h0;

(*@\mutantid{M7}@*) assign out_and =
  &{in[99:65], 1'h1, in[63:0]};
\end{lstlisting}

\vspace{-6pt}

\mutheading{OR reduction faults}

\begin{lstlisting}[style=verilogstyle,escapeinside={(*@}{@*)}]
(*@\mutantid{M1}@*) assign out_or = 1'h0;

(*@\mutantid{M6}@*) assign out_or =
  |{in[99:14], 1'h0, in[12:0]};
\end{lstlisting}

\vspace{-6pt}

\mutheading{XOR reduction faults}

\begin{lstlisting}[style=verilogstyle,escapeinside={(*@}{@*)}]
(*@\mutantid{M3}@*) assign out_xor =
  ^{in[99:67], ~in[66], in[65:0]};

(*@\mutantid{M5}@*) assign out_xor = 1'h1;

(*@\mutantid{M8}@*) assign out_xor = 1'h0;

(*@\mutantid{M9}@*) assign out_xor = ~(^in);
\end{lstlisting}

\vspace{1pt}

{\tiny\color{black!65}
CovR kills M0, M4, and M6 that Pro-V misses; M7 survives both.
}

\end{tcolorbox}
\end{minipage}

\vspace{-4pt}

\begin{tcolorbox}[title={(d) Pro-V Stimulus}]
\centering
\scriptsize
\textbf{Mutation Score:} 60.0\% \qquad
\textbf{Killed:} 6/10

\vspace{2pt}

{\fontsize{5.8}{6.3}\selectfont
\setlength{\tabcolsep}{5pt}
\renewcommand{\arraystretch}{0.96}
\begin{tabular}{c|ccccc}
\textbf{Test} & 0 & 1 & 2 & 3 & 4 \\
\hline
\texttt{in} &
\texttt{0011...a6} &
\texttt{1110...4d} &
\texttt{1110...e2} &
\texttt{0010...86} &
\texttt{0011...9f} \\
\texttt{out\_and} & 0 & 0 & 0 & 0 & 0 \\
\texttt{out\_or}  & 1 & 1 & 1 & 1 & 1 \\
\texttt{out\_xor} & 1 & 0 & 0 & 1 & 0 \\
\end{tabular}
}

\vspace{3pt}

\raggedright
\scriptsize
\textbf{Killed Mutants:}
\killchip{M1} \killchip{M2} \killchip{M3}
\killchip{M5} \killchip{M8} \killchip{M9}

\textbf{Survived Mutants:}
\survivechip{M0} \survivechip{M4}
\survivechip{M6} \survivechip{M7}

\vspace{2pt}
\footnotesize
Random dense vectors usually have \texttt{out\_and=0} and
\texttt{out\_or=1}. This misses mutants that require all-one, all-zero, or
specific one-hot patterns.
\end{tcolorbox}

\vspace{-3pt}

\begin{tcolorbox}[title={(e) CovR Stimulus}]
\centering
\scriptsize
\textbf{Mutation Score:} 90.0\% \qquad
\textbf{Killed:} 9/10

\vspace{2pt}

{\fontsize{6.2}{6.8}\selectfont
\setlength{\tabcolsep}{5pt}
\renewcommand{\arraystretch}{0.96}
\begin{tabular}{c|ccccc}
\textbf{Test} & 0 & 1 & 2 & 17 & 1206 \\
\hline
\texttt{in} &
\texttt{0...0} &
\texttt{0...1} &
\texttt{0101...01} &
\texttt{bit13=1} &
\texttt{1...1} \\
\texttt{out\_and} &
0 & 0 & 0 & 0 &
\textcolor{covrgreen}{\textbf{1}} \\
\texttt{out\_or} &
\textcolor{covrgreen}{\textbf{0}} &
1 & 1 &
\textcolor{covrgreen}{\textbf{1}} &
1 \\
\texttt{out\_xor} &
0 & 1 & 0 & 1 & 0 \\
\end{tabular}
}

\vspace{3pt}

\raggedright
\scriptsize
\textbf{Killed Mutants:}
\killchip{M0} \killchip{M1} \killchip{M2} \killchip{M3}
\killchip{M4} \killchip{M5} \killchip{M6}
\killchip{M8} \killchip{M9}

\textbf{Survived Mutants:}
\survivechip{M7}

\vspace{2pt}
\footnotesize
CovR includes reduction-specific edge cases: all-zero for \texttt{out\_or},
all-one for \texttt{out\_and}, alternating patterns for parity, and one-hot
vectors for bit-sensitivity faults.
\end{tcolorbox}

\vspace{-3pt}

\begin{tcolorbox}[
  title={(f) Mutants Newly Killed by CovR},
  colback=black!1,
  colframe=black!75
]
\centering
\scriptsize
\setlength{\tabcolsep}{4pt}
\renewcommand{\arraystretch}{1.08}

\resizebox{\linewidth}{!}{%
\begin{tabular}{c|l|c|c|c}
\toprule
\textbf{Mutant} &
\textbf{Fault} &
\textbf{Killing stimulus} &
\textbf{Expected} &
\textbf{Mutant output} \\
\midrule

M0 &
\texttt{out\_and} uses \texttt{\string~in[76]} &
\texttt{in = 1...1} &
\texttt{out\_and = 1} &
\textcolor{red!75!black}{\texttt{out\_and = 0}} \\

M4 &
\texttt{out\_and = 1'h0} &
\texttt{in = 1...1} &
\texttt{out\_and = 1} &
\textcolor{red!75!black}{\texttt{out\_and = 0}} \\

M6 &
\texttt{out\_or} ignores bit 13 &
\texttt{in[13] = 1}, others 0 &
\texttt{out\_or = 1} &
\textcolor{red!75!black}{\texttt{out\_or = 0}} \\

\bottomrule
\end{tabular}%
}

\vspace{2pt}

\raggedright
\footnotesize
The three additional kills come from structured vectors that directly target
reduction-operator corner cases rather than relying on random dense inputs.

\end{tcolorbox}

\caption{
CovR improves mutation detection on \texttt{Prob052\_gates100}. Pro-V's
random vectors kill parity and obvious stuck-output mutants, but miss
reduction-specific corner cases. CovR adds all-zero, all-one, alternating,
and one-hot stimuli, improving mutation score from 6/10 to 9/10.
}
\label{fig:covr-mutant-killing-gates100}
\end{figure}

\end{appendix}

\end{document}